\documentclass[twocolumn]{aastex631}
\usepackage{graphicx}	
\usepackage{grffile}

\usepackage{bm}

\newcommand{\m}{\ensuremath{\,{\rm m}}}
\newcommand{\km}{\ensuremath{\,{\rm km}}}
\newcommand{\K}{\ensuremath{\, {\rm K}}}

\newcommand{\MHz}{\ensuremath{\, {\rm MHz}}}
\newcommand{\Jy}{\ensuremath{\,{\rm Jy}}}
\newcommand{\mJy}{\ensuremath{\,{\rm mJy}}}

\renewcommand{\deg}{\ensuremath{\,{\rm deg}}}
\renewcommand{\arcmin}{\ensuremath{\,{\rm arcmin}}}
\renewcommand{\sec}{\ensuremath{\,{\rm s}}}
\newcommand{\nside}{\ensuremath{\,{N_{\rm side}}}}

\newcommand{\vecn}{\ensuremath{{\hat{\mathbf{n}}}}}
\newcommand{\vecr}{\ensuremath{{\mathbf{r}}}}
\newcommand{\vecv}{\ensuremath{{\mathbf{v}}}}

\newcommand{\pix}{\ensuremath{{\rm pix}}}
\newcommand{\moon}{\ensuremath{{\rm moon}}}
\newcommand{\matV}{\ensuremath{{\mathbf{V}}}}
\newcommand{\matB}{\ensuremath{{\mathbf{B}}}}
\newcommand{\matN}{\ensuremath{{\mathbf{N}}}}
\newcommand{\matR}{\ensuremath{{\mathbf{R}}}}
\newcommand{\matQ}{\ensuremath{{\mathbf{Q}}}}
\newcommand{\matW}{\ensuremath{{\mathbf{W}}}}
\newcommand{\matI}{\ensuremath{{\mathbf{I}}}}

\newcommand{\matSigma}{\ensuremath{{\mathbf{\Sigma}}}}
\newcommand{\mats}{\ensuremath{{\mathbf{T}}}}

\newcommand{\matsnew}{\ensuremath{{\hat{\mathbf{T}}}}}
\newcommand{\BTB}{\ensuremath{{ \mathbf{B}^T\mathbf{N}^{-1} \mathbf{B}} }}
\newcommand{\BTV}{\ensuremath{{ \mathbf{B}^T\mathbf{N}^{-1} \mathbf{V}} }}
\newcommand{\BT}{\ensuremath{{ \mathbf{B}^T\mathbf{N}^{-1}} }}

\newcommand{\snrl}{\ensuremath{ {\rm SNR}_l}}
\shorttitle{DSL}
\shortauthors{Deng et al.}

\begin{document}

\title{Synthesis imaging with a lunar orbit array: I. Global Sky Map and its Systematics }

\correspondingauthor{Furen Deng; Xuelei Chen; Yidong Xu}
\email{frdeng@nao.cas.cn; xuelei@cosmology.bao.ac.cn; xuyd@nao.cas.cn}

\author[0000-0001-8075-0909]{Furen Deng}
\affiliation{National Astronomical Observatories, Chinese Academy of Sciences, 20A Datun Road, Chaoyang District, Beijing 100101, China}
\affiliation{School of Astronomy and Space Science, University of Chinese Academy of Sciences, Huairou District, Beijing 101408, China}
\affiliation{State Key Laboratory of Radio Astronomy and Technology, Beijing 100101, China}

\author[0000-0003-3224-4125]{Yidong Xu}
\affiliation{National Astronomical Observatories, Chinese Academy of Sciences, 20A Datun Road, Chaoyang District, Beijing 100101, China}
\affiliation{State Key Laboratory of Radio Astronomy and Technology, Beijing 100101, China}

\author[0000-0002-6174-8640]{Fengquan Wu}
\affiliation{National Astronomical Observatories, Chinese Academy of Sciences, 20A Datun Road, Chaoyang District, Beijing 100101, China}
\affiliation{State Key Laboratory of Radio Astronomy and Technology, Beijing 100101, China}

\author[0000-0002-4456-6458]{Yanping Cong}
\affiliation{Shanghai Astronomical Observatory, Chinese Academy of Sciences, 80 Nandan Road, Xuhui District, 200030, China}

\author[0000-0002-7829-1181]{Bin Yue}
\affiliation{National Astronomical Observatories, Chinese Academy of Sciences, 20A Datun Road, Chaoyang District, Beijing 100101, China}
\affiliation{School of Astronomy and Space Science, University of Chinese Academy of Sciences, Huairou District, Beijing 101408, China}
\affiliation{State Key Laboratory of Radio Astronomy and Technology, Beijing 100101, China}

\author[0000-0001-6475-8863]{Xuelei Chen}
\affiliation{National Astronomical Observatories, Chinese Academy of Sciences, 20A Datun Road, Chaoyang District, Beijing 100101, China}
\affiliation{School of Astronomy and Space Science, University of Chinese Academy of Sciences,  Huairou District, Beijing 101408, China}
\affiliation{State Key Laboratory of Radio Astronomy and Technology, Beijing 100101, China}


\begin{abstract}
Ground-based radio astronomical observation at frequencies below 30 MHz is hampered by the Ionosphere and radio frequency interference (RFI). The Discovering Sky at the Longest wavelength (DSL) mission, also known as the Hongmeng mission, employs a linear array of satellites on a circular orbit around the Moon to make interferometric observations in this band. Though vastly different from the usual ground-based arrays, the interferometric visibility data collected by such an array is linearly related to the sky map, and the reconstruction is in principle an inversion problem of linear mapping.  In this paper, we investigate a number of issues in the algorithm of global map reconstruction, focusing on the impact of sub-pixel noise induced by the finite pixelization of the sky, and errors due to regularization. We find that in the reconstruction process, if one builds up the beam matrix, which relates the sky pixels to the visibilities, by naively evaluating its elements at each of the pixel centers, then the sub-pixel noise can give rise to a significant aliasing effect. However, this effect can be effectively mitigated by a simple pixel-averaging method. Based on evaluation of the image quality using the correlation coefficient between the input and reconstructed map, and the signal-to-noise ratio, we discuss the selection strategy of the regularization parameter, and show that the sky can be well reconstructed with a reasonable choice of the regularization parameter.
\end{abstract}


\section{Introduction} \label{sec:intro}
Radio astronomical observations at frequency $\lesssim 30\MHz$ are important for the detection of 21 cm signals from the Dark Ages \citep{2023NatAs...7.1025M}, the structure of the Milky Way \citep{2022ApJ...940..180C}, and for the evolution of radio sources \citep{2022A&A...668A.186S}, but they are severely impeded by the absorption and reflection of the ionosphere and radio frequency interferences (RFIs). The far-side of the lunar surface and orbit is free from a thick ionosphere, and the Moon can be used to block the interference from Earth, therefore ideal for making such low-frequency observations \citep{Chen2019}. Pioneering low frequency radio astronomical experiments have been carried out during the Chang'e-4 mission, including the lunar lander low frequency observations \citep{Zhu2021}, the Netherlands-China Low-frequency Explorer (NCLE) experiment on board the Queqiao relay satellite \citep{Vecchio2021}, and the Longjiang microsatellite experiments \citep{2023ExA....56..333Y}
; and the ROLSES-1 experiment on board the  Odysseus lunar lander \citep{Hibbard2025}. A number of mission concepts have also been proposed, some recent lunar surface mission concepts include the FARSIDE \citep{2021RSPTA.37990564B}, Lunar Crater Radio Telescope (LCRT, \citealt{LCRT2021}), Lunar Surface Electromagnetics Experiment-Night (LuSEE-Night, \citealt{2023arXiv230110345B}), the Astronomical Lunar Observatory (ALO, \citealt{2024AAS...24326401K}), and the Large-scale array for radio astronomy on the farside (LARAF, \citealt{chen2024large}); and lunar orbit missions include the single satellite DARE and DAPPER \citep{2017arXiv170200286P}, PRATUSH \citep{2023ExA....56..741S} and SEAMS \citep{SEAMS2021}, CosmoCube \citep{Artuc2024, zhu2025}, and the array Discovering Sky at the Longest wavelength (DSL) \citep{Chen:2020lok,chen2023}. Missions on the surface of the Moon face engineering challenges such as construction, power supply and data transmission, and the charged dust near the lunar surface may also affect radio observation. Lunar orbit missions are simpler from an engineering perspective, but its data may require more sophisticated processing, as there are a number of differences for the imaging observation of such an orbital array from the usual ground-based interferometric observation.  

The DSL satellite array consists of a mother satellite, which handles data communication, and 9 daughter satellites for astronomical observation \citep{Chen:2020lok,chen2023}. One daughter satellite is dedicated to the precise measurement of the global spectrum in the $30-120$ MHz band \citep{Shi:2022zdx,wu2024}, and the other eight daughter satellites form a linear array to make interferometric imaging observation and global spectrum measurement in the $0.1 - 30$ MHz band. Each of these eight satellites carry three pairs of orthogonal short dipole antennas, which are electrically small, i.e. the size of the antenna is much shorter than the wavelength, so the beam pattern of each pair of dipole antenna is that of a short dipole, which is almost omnidirectional. 

Astronomical observations are carried out by the daughter satellites when they are on the far side part of the orbit, where the radio frequency interference (RFI) from Earth is blocked by the Moon. Interferometry data from each daughter satellite are transmitted to the mother satellite  via a microwave link at the much higher frequency of the Ka band. Due to the limit of the inter-satellite communication bandwidth, only a selection of 30 channels of 8 kHz channel width within the 0.1-30 MHz band are transmitted.  Cross-correlations are made on the mother satellite to produce the visibility data. 
When the satellite array is on the near side of the orbit where the Earth is in sight, the mother satellite transmits the visibility and other auxiliary data, such as the baseline vectors, orientations of the satellites, and calibration measurements to ground stations. The calibration and imaging processes are carried out offline. For more details, we refer readers to \cite{Chen:2020lok} and \cite{chen2023}.

As the satellites move along their orbits, and with the precession of the orbital plane, they generate baselines which are distributed in three dimension, though for each baseline a part of the sky is blocked by the Moon. The synthesis of all-sky image from such data is the topic of the present study. Many algorithms had been proposed for wide-field imaging of ground based arrays, such as faceting \citep{1992A&A...261..353C}, w-projection \citep{2008ISTSP...2..647C} and w-stacking \citep{2012SPIE.8500E..0LC}. However, these algorithms are mainly developed for the `snapshot' observation of sky patches, not particularly suited for the completely 3D distribution of baselines as generated by the satellite motions. These methods have also not taken into account the position-dependent blocking of the sky by the Moon. However, we can view the visibility data obtained by the satellite array as a linear map from the sky intensity, with complicated but known mapping relation. The synthesis imaging of the sky can then be viewed as a linear inversion problem, and can in principle be solved, as demonstrated in the noise-free case \citep{Huang2018} and with noise \citep{Shi:2022xdw}. 

However, this algorithm still have a number of systematic issues when applied in practice. Specifically, as is often encountered in astronomical synthesis imaging, the  $uvw$-coverage may be incomplete, so the solution of the inversion problem is not unique, and generally requires taking a regularization method, which depends on the nature and magnitude of the noise, and results in an imperfect ``effective beam''.  Furthermore, there are also other systematic effects which can degrade the reconstructed map. In particular, solution of the inversion problem requires a pixelization of the sky of finite resolution, but the noise on the sub-pixel scales may generate significant distortions if not properly treated. In this work, we study these effects by numerical simulation, focusing on the reconstruction of the full-sky map. We quantify the effects of imperfect ``effective beam'', thermal noise, and the sub-pixel noise, and discuss trade-offs in the choice of regularization parameter in imaging reconstruction.

Besides the imaging algorithm, there are also other sources of systematics which may distort or degrade the synthesized image, such as the errors in the instrumental phase and gain calibration, inaccurate measurement of positions and orientations of satellites, and the error in clock synchronization between satellites. In the DSL mission, the orientations of each satellite, and the direction of each daughter satellite with respect to the mother satellite which is equipped with light signal are determined by star sensors on the daughter satellite; the clock synchronization and distance measurement between satellites are achieved by the microwave link between the daughter satellites and the mother satellite. Additionally, an artificial calibration signal is used to help determine the phase and amplitude of the receiver gain. In the present work we will focus on the imaging algorithm and neglect these systematic errors, their impact will be studied in a subsequent paper (M. Zhou et al., {\it in preparation}) in more details, but the conclusions presented here will be robust against such systematics, provided the performance of the satellite array attains the design parameters.

This paper is organized as follows.  We introduce a sky model with structures on all relevant scales in Section \ref{sec:mock_sky}, which is needed to properly assess the effect of sub-pixel noise.  Then we describe our simulation set up in Section \ref{sec:setup}, including the satellite array configuration, orbital motion, and the `breathing' of the satellite array. We model the received signal and noise, and discuss the effect of  finite time resolution and frequency bandwidth, and the effect of lunar reflection. Then, in  Section \ref{sec:map}, we introduce the formalism of sky map reconstruction from the visibility data with the Tikhonov regularization. There is a critical problem in this reconstruction: the aliasing effect due to the finite resolution of pixelization. We show that with sub-pixel noise, this can pose a serious problem, but it can be effectively mitigated by applying a simple pixel-averaging method, and present the reconstruction results for a number of setups.  In Section \ref{sec:error}, we consider  various errors in the image reconstruction, including the thermal noise, as well as the beam error induced by the aliasing effect, and we study their impacts on the reconstruction, and discuss the choice of the Tikhonov regularization parameter to minimize the distortion due to these various effects. Finally, we summarize the results and conclusion in Section \ref{sec:conclusion}.

\section{Sky Map with Small Scale Power} \label{sec:mock_sky}
In previous works on the image reconstruction for the lunar orbit array, the simulation were performed using readily available sky models as inputs, such as the improved global sky map (GSM) model \citep{deOliveiraCosta:2008pb,Zheng2017} in \citet{Huang2018}, and the self-consistent whole sky foreground model (SSM) \citep{Huang2019} in \citet{Shi:2022xdw}. However, due to their limited resolution, these sky models are lacking in anisotropy power on small scales. 
We need a full sky map with relatively high angular resolution as input for this simulation. 

\begin{figure}
\gridline{
\fig{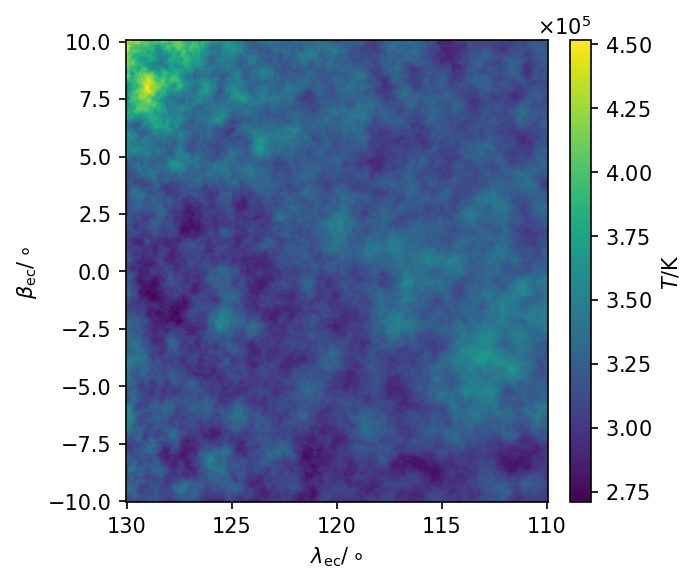}{0.45\textwidth}{(a) A sky patch for the diffuse component.}
}
\gridline{
\fig{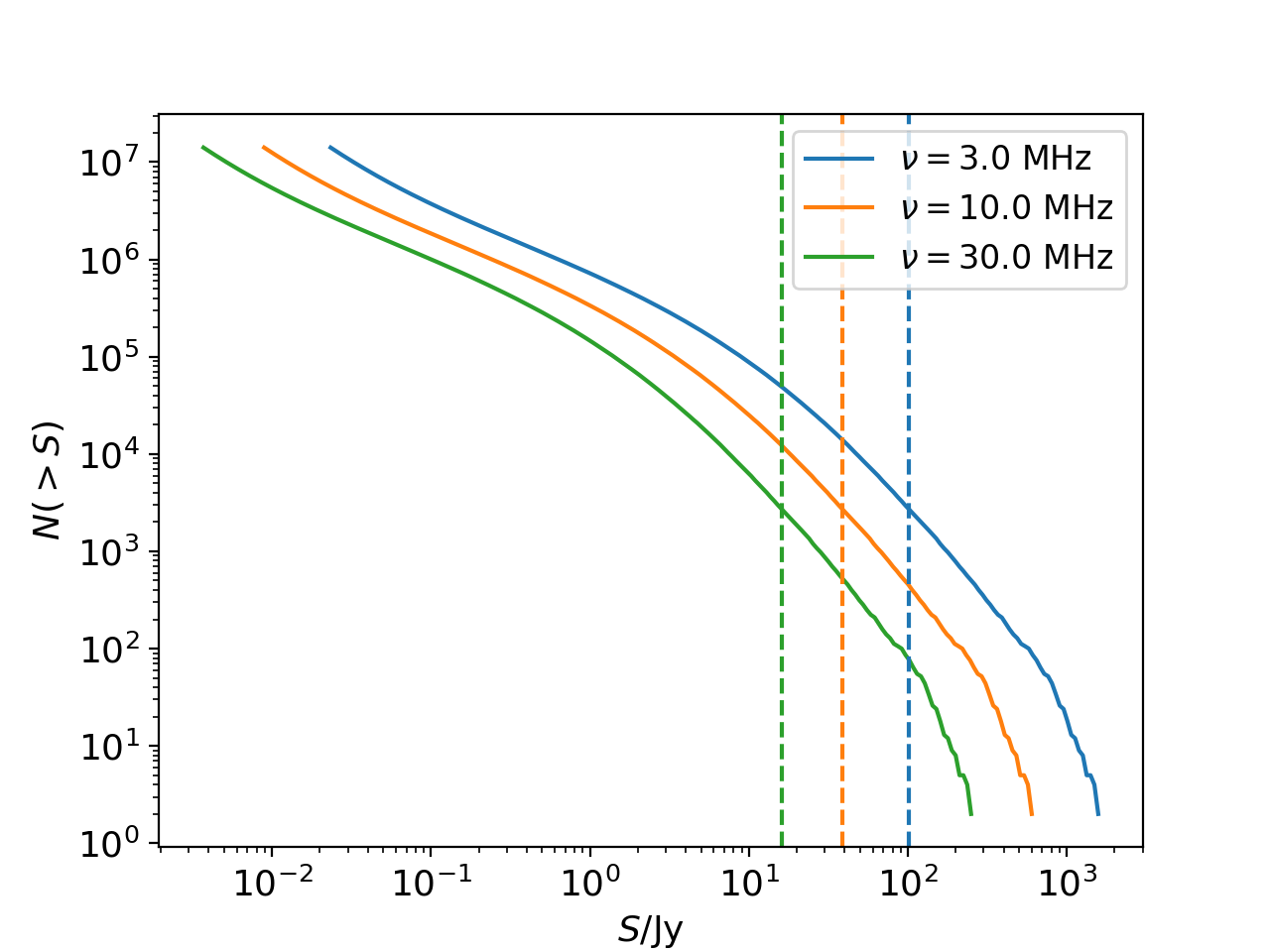}{0.45\textwidth}{(b) The cumulative flux distribution for the point sources.}
}
\caption{The input sky model. (a) A patch of the sky of the diffuse component at 10 MHz, shown in the ecliptic coordinates ($\lambda_{ec}, \beta_{ec}$).  The location of this patch is marked by the lowest red rectangle in Figure \ref{fig:full_sky_map}. (b) Cumulative flux distribution of the point sources as a function of the flux $S$ in units of Jy} at 3, 10 and 30 MHz, respectively, the low-flux cut-off (see text) is indicated by the dashed vertical lines. 
\label{fig:lum_skymap}
\end{figure}

As existing full sky maps do not have sufficient resolution, we construct a mock map, by including the Galactic diffuse emission and point sources separately. 
We adopt the direction dependent spectral index model of the Ultra Long wavelength Sky model with Absorption (ULSA) \citep{2021ApJ...914..128C}, which provides the Galactic diffuse emission sky map (point sources subtracted) for low frequencies down to $\sim 1\MHz$ with the 
Hierarchical Equal Area iso-Latitude Pixelization (\texttt{HEALPix}) scheme \citep{2005ApJ...622..759G,Zonca2019} 
as the base map for our construction. The HEALPix scheme is frequently used in the pixelization of the full sky or other spherical surface. It starts with the division of the sphere into 12 equal-area facets, and then hierarchically divides each facet into $\left(\nside\right)^2$ equal-area pixels, so that the whole sky is divided into $12 \left(\nside\right)^2 $ equal-area pixels. 
The ULSA model is based on the updated Haslam map \citep{2015MNRAS.451.4311R}, its angular resolution is limited by that of the Haslam map which is $\sim 1^\circ$, insufficient to assess our synthesis imaging algorithm on smaller scales. We substitute the low-resolution Haslam map used by ULSA with a high-resolution version, where the small-scale fluctuations are added following \cite{2015MNRAS.451.4311R} (See their Section 4.2). 
We adopt the ecliptic coordinates for maps in this paper, as both the orbital plane of the Moon around Earth and the equatorial plane of the Moon are close to the ecliptic plane. We use $\lambda_{\rm ec}$ and $\beta_{\rm ec}$ to denote the longitude and latitude in the ecliptic coordinates. 
Figure \ref{fig:lum_skymap}(a) shows a $20\deg\times 20\deg$ patch of the mock diffuse map at 3 MHz, centered at $\lambda_{\rm ec}=120^\circ, \beta_{\rm ec}=0^\circ$, with the small-scale fluctuations included.

Mock point sources are randomly generated, according to a distribution derived from the source counts of the GaLactic and Extragalactic All-sky MWA survey (GLEAM) at 200, 154, 118, and 88 MHz \citep{2019PASA...36....4F}. The low and high flux cut-off for the GLEAM model at 154 MHz are $1\mJy$ and $75\Jy$ respectively \citep{2019PASA...36....4F}. However, the point sources with flux $\lesssim 2\Jy$ are not removed in the stripping process for the Haslam map, and they contribute to the diffuse component \citep{2015MNRAS.451.4311R}. 
To avoid double counting, we extrapolate the confusion limit of Haslam map (2 Jy) at 408 MHz to 154 MHz, which is 4.36 Jy for a spectral index of 0.8 \citep{2019PASA...36....4F}. It is well above the GLEAM lower cut-off, so we only keep the point sources with flux above 4.36 Jy at 154 MHz. The cumulative flux distribution of the simulated sources at 3, 10 and 30 MHz are shown in Figure \ref{fig:lum_skymap} (b), and the low-flux cutoffs are indicated by the dashed lines. The remaining bright point sources contribute $\sim 70\%$ of the angular power spectrum of the point sources, so the small-scale power contributed by point sources is not severely affected by the flux cut-off.

\begin{figure}
\fig{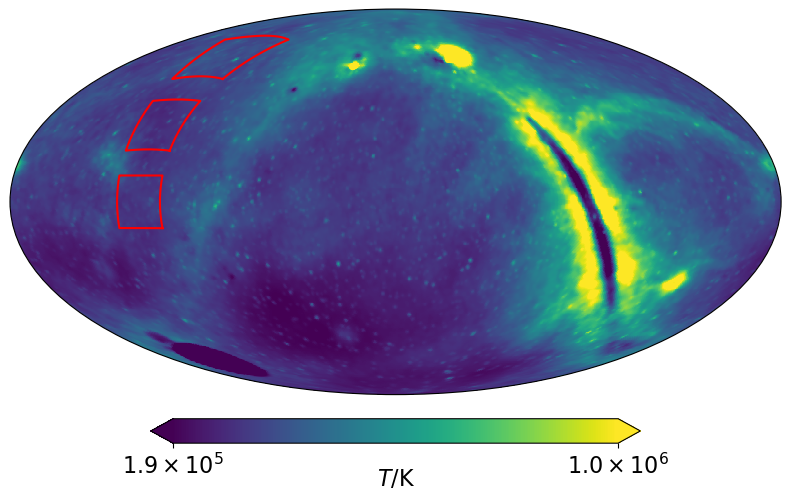}{0.5\textwidth}{}
\caption{The full-sky map with both diffuse component and point sources in ecliptic coordinates at 10 MHz. The map is smoothed by a Gaussian filter with Full-Width-Half-Maximum (FWHM)$\approx 1.3^\circ$ to match the resolution of our reconstructed map. The red boxes mark small regions where we take detailed study. 
}
\label{fig:full_sky_map}
\end{figure}

Both diffuse and point-source maps are first generated with $\nside=2048$ and then downgraded to $\nside=1024$.  The full-sky map with both diffuse component and point sources at 10 MHz is shown in Figure \ref{fig:full_sky_map}. 
The Galactic plane is inclined at an angle of about $60^\circ$ to the ecliptic plane. Note that at such low frequencies, the free-free absorption of the free electrons in the interstellar medium (ISM) is quite strong, so the center of the galactic plane actually becomes a dark belt due to the large density of the ISM near the Galactic plane. However, the so called North Polar Spur (appearing as an arc near the right edge of this map) is still bright, as it is modeled here as the emission of a nearby structure in the ULSA model \citep{2021ApJ...914..128C}, though if it is located further away it may also be darkened \citep{cong2024loop}. The dark blob at the bottom left of the plot is the Gum nebula, which is opaque at this low frequency.

\section{Simulation Setup}
\label{sec:setup}
\subsection{Satellite array configuration and orbital motion}\label{sec:orbit_bls}
The array of satellites rely on the orbital motion and precession of the orbital plane to achieve 3D $uvw$ baseline coverage \citep{Huang2018, Shi:2022xdw}. In this work we follow the same simplifying assumption that all satellites moves on the same orbit without deviation.
We model the orbital motion of the satellite array as a combination of a uniform circular motion at a height of 300 km and an inclination angle of $30^\circ$ with an orbital period of 8248.68 s, and a uniform precession of orbital plane with a period of 1.3 year \citep{Shi:2022xdw}. 
We assume that the observation is only made when the Earth is shaded by the Moon, which is $\sim 1/3$ of the total orbital time to avoid the RFI from the Earth.

We shall refer the 8 daughter satellites for interferometric imaging by ID number 1, 2, ...8, with number 1 being the closest to the mother satellite. As in many ground-based arrays, the spacings of the satellite array are non-uniform, with a wide range distribution to produce both short and long baselines. This spacing configuration can be optimized to produce better maps, but it is also subject to various practical constraints. Here as a simple model to generate baselines of different scales, we assume the satellite positions satisfy a power law form with an offset, 
\begin{eqnarray}\label{eq:pos_ini}
    r^0_n = a_{\rm offset} + a_0 q_0^{n-2} \quad {\rm for}\ n\geq 2,
\end{eqnarray}
where $r_n^0$ is the initial position of the daughter satellite $n$, and take the first daughter satellite's position as the reference point, i.e. $r^0_1=0$. As $n$ goes from 2 to 8, the daughter satellites are successively away from the first daughter satellite, and therefore $q_0>1$ and $a_0>0$.

To achieve a better $uvw$ coverage in a shorter time scale to accommodate the diversity of science cases, and to correct for naturally generated small velocity differences between the satellites, a formation reconfiguration strategy which we shall refer as {\it breathing} can be implemented. The satellites are commanded to move slowly towards or away from each other periodically, so as to cover different baseline lengths. This can be achieved by adding a slightly different relative velocity for each satellite. 
One breathing period is divided into 2 phases with equal duration. 
Starting from the initial configuration specified by Eq.~(\ref{eq:pos_ini}), the array first compresses towards satellite 1 in the first phase, and then stretches back to the initial configuration in the second phase. 

In the present work, we shall adopt a simplified assumption that the satellites are moving with constant speed for most time and undergo instant velocity change at the beginning of each phase of the breathing motion. The breathing motion is then completely specified by the array configuration at the beginning and the end of the first phase, and the breathing period. We link these two configurations by the compression ratio $\mathcal{R}$, which is the ratio of the baselines in the two configurations, i.e. the amplitude of the breathing motion. 
Note that in the actual case if the relative movements between satellites are different from this specification, the exact $uvw$ distribution would be slightly different, but the overall coverage and density would still be similar.
Based on considerations of orbital dynamics and engineering practicality, here we adopt a breathing period of 14 days, though this may be adjusted to fit the engineering requirement, and would not affect the result much. In Table \ref{tab:breath_pos}, we list the positions of the satellites at the beginning of the cycle of 2 phases, with respect to the position of the first satellite at the beginning of each breathing period. For a more flexible breathing scheme, we allow the compression ratio $\mathcal{R}$ to vary between different satellites and use subscript $n$ to indicate different satellites. We notice that $\mathcal{R}_1$ does not have any effect because $r_1^0=0$ by definition.

\begin{table}
\caption{The positions of satellites at the beginning of the 2 phases, with respect to the initial position of the first satellite, where $r^0_n$ are the positions at the beginning of each breathing period and is given by Eq.~(\ref{eq:pos_ini}), and $\mathcal{R}_n$ are the compression ratios which we allow to vary between different satellites. }
\label{tab:breath_pos}
\begin{center}
\begin{tabular}{c|c}
\hline
  Time (day)   &  $r_n$ \\
  \hline
   0   & $r_n^0$ \\
   7   & $r_n^0/\mathcal{R}_n$ \\
   14  & $r_n^0$ \\
\hline   
\end{tabular}    
\end{center}
\end{table}

\begin{figure}
\fig{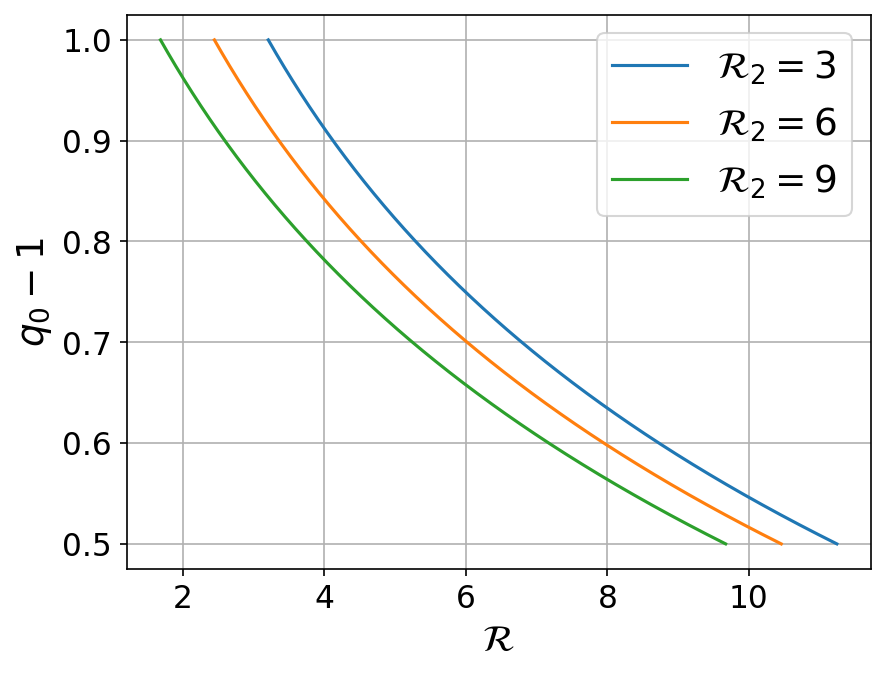}{0.45\textwidth}{}
\caption{The geometric factor $q_0$ as a function of the compression ratio $\mathcal{R}$ for the satellites configuration.
The blue, orange, and green lines correspond to the compression ratio  between satellites 1 and 2  of $\mathcal{R}_2 = $3, 6, and 9, respectively.}\label{fig:geofactor}
\end{figure}

To avoid the risk of collision, the  shortest baseline is required to be $ \geq 100$ m, and the second shortest baseline is required to be $\geq 500$ m. The longest baseline is chosen to be 100 km. Longer baselines can in principle introduce higher angular resolution, but would require higher power for communication and better angular position determination to achieve sufficient precision in baseline determination. At the same time, the angular resolution of the array is also limited by the the angular broadening effect caused by scattering in the interstellar medium and the interplanetary medium. A maximum baseline of 100 km is chosen as a balance of these considerations \citep{Chen:2020lok}. To satisfy these constraints, 
for a given set of compression ratio $\mathcal{R}_n$, the parameters $a_{\rm offset}$, $a_0$ and $q_0$ are determined by imposing the requirements on the shortest and the longest distance between satellites. The positions of daughter satellites are measured from the mother satellite, and its measurement error will be larger for the more distant satellites. To reduce the impact of the positional error, we set the shortest baseline to be the one between daughter satellites 1 and 2, and set the second shortest baseline to be the one between satellites 2 and 3. The spacing of the satellites increases with the satellite ID number, and the longest baseline is reached between the daughter satellites 1 and 8. These constraints yield the following equations:
\begin{eqnarray}
r^0_2 &=& \mathcal{R}_2\times 100\m,\label{eq:bl_req_1} \\
r^0_3-r^0_2 &=& \mathcal{R}_3 \times 500\m,\label{eq:bl_req_2}\\
r^0_8 &=& 100\km\label{eq:bl_req_3}.
\end{eqnarray}
The simplest choice is a universal compression ratio $\mathcal{R}$ for all satellites. However, it is also possible to adopt different comparison ratio for the different satellites. In particular, we note that if $\mathcal{R}_2<5$, there would be a gap in the baseline length distribution due to the constraints Eq.~(\ref{eq:bl_req_1}) and (\ref{eq:bl_req_2}). We therefore vary $\mathcal{R}_2$ independently and set $\mathcal{R}_n=\mathcal{R}$ for $n\neq 2$. 
Nevertheless we would show later that the baseline length distribution is not sensitive to $\mathcal{R}$. The geometric factor $q_0$ for different $\mathcal{R}_2$ and $\mathcal{R}$ is shown in Figure \ref{fig:geofactor}. 
For larger compression ratios $\mathcal{R}_2$ and $\mathcal{R}$, $q_0$ is constrained to be closer to 1, and the satellite array becomes closer to an evenly spaced array, while for smaller compression ratios different scales are sampled at an instant. However, note that due to breathing motion, even for uniformly spaced array configuration, different baseline lengths would be generated during the course of operation.

\begin{figure}
\gridline{
\fig{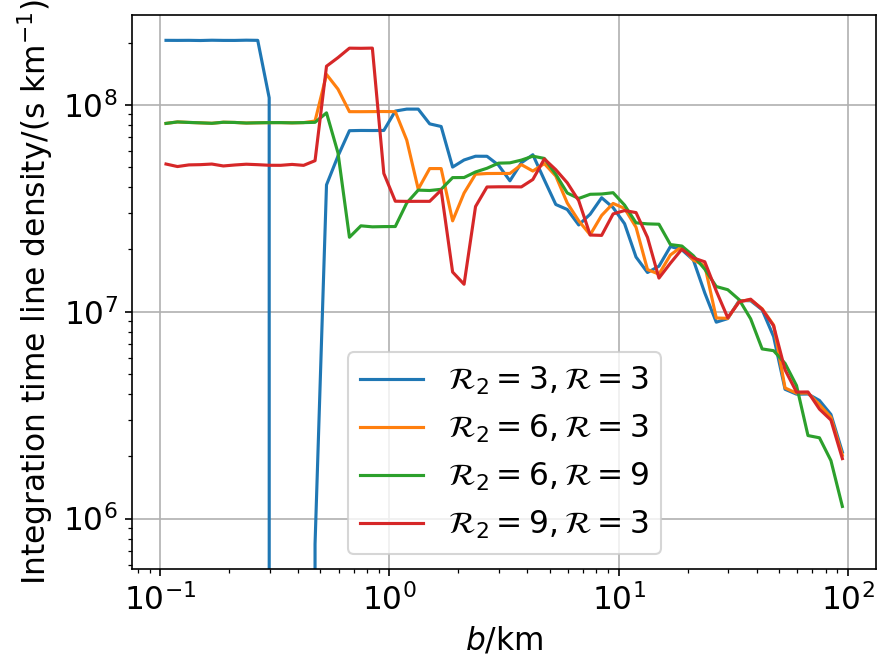}{0.45\textwidth}{}
}
\caption{The integration time spent per unit baseline length as a function of baseline length $b$ for different compression ratios $\mathcal{R}_2$ and $\mathcal{R}$ of the `breathing' of the satellite array, after one cycle of precession (1.3 years).
}\label{fig:baseline_1d}
\end{figure}

\begin{figure}
\gridline{
\fig{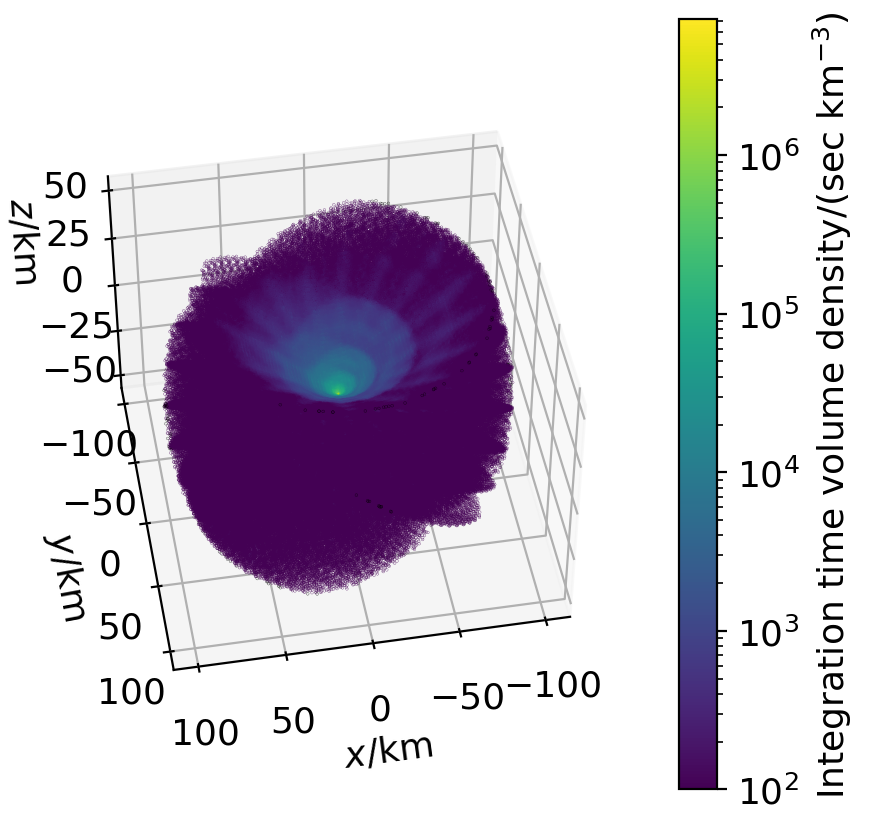}{0.5\textwidth}{}
}
\caption{The 3D distribution of the integration time volume density of baselines of the DSL array, generated after one cycle of precession(1.3 years), for the $\mathcal{R}_2=6$, $\mathcal{R}=3$ case.}\label{fig:baseline_3d}
\end{figure}

The integration time spent per unit baseline length as a function of baseline length $b$ after one cycle of precession (1.3 years) for different compression ratios $\mathcal{R}_2$ and $\mathcal{R}$ is shown in Figure \ref{fig:baseline_1d}.
There is a sharp drop between $100\m < b <500 \m$ for the $\mathcal{R}_2=3$, $\mathcal{R}=3$ case, because there is the constraint that the shortest distance 100 m and the second shortest distance 500 m and for this case $r_2^0=\mathcal{R}_2\times 100\m < 500 \m$. Once $\mathcal{R}_2>5$, further increasing $\mathcal{R}_2$ does not change the distribution much. The distribution is insensitive to $\mathcal{R}$ which mainly affect long baselines, indicating that the baseline length distribution at long baseline end is insensitive to the breathing scheme.

For the larger compression ratio cases,  during the breathing this range of baseline length is covered as two satellites come close, but this is not intended because at any given moment such short spacing should be avoided as a risk of collision, and if the short baseline is needed, they should be generated by the two satellites nearest to the mother satellite, which are the best monitored ones in the array. Therefore we choose $\mathcal{R}_2=6$ and $\mathcal{R}=3$ as our fiducial case which could produce more uniform baseline length distribution at short baseline end while reducing the risk of collision.

The integration time spent per unit volume for our fiducial configuration with `breathing' is shown in Figure \ref{fig:baseline_3d}. The sandglass-like shape is produced by the rotation of an inclined plane distribution of the baselines. 
In this paper, we consider the observation time of 1.3 yr, corresponding to one full cycle of precession.

\subsection{Signal and Thermal Noise}

We model the Moon as an opaque sphere with a radius of $R_\moon=1737.1\km$. The radio emission of the Moon itself is mainly thermal, which is typically much smaller than the sky brightness at this low frequency \citep{Zhang_2012} and can be neglected. The Moon does reflect radio waves, but this can also be neglected (see Section \ref{sec:reflection} for more discussions). The visibility of baseline $ij$ is given by
\begin{equation}\label{eq:vis}
V_{ij} = \int T(\vecn) S_{ij}(\vecn) A_{ij}(\vecn) \exp\left(-2\pi i\frac{\nu}{c} \vecn\cdot \vecr_{ij}\right) d\vecn,
\end{equation}
where $T(\vecn)$ is the sky map, $S_{ij}(\vecn)=S_i(\vecn) S_j(\vecn)$ is the shading by the moon, $A_{ij}(\vecn) = \sqrt{A_i(\vecn) A_j(\vecn)}$ is the power response of the primary beam and $\vecr_{ij}=\vecr_i - \vecr_j$ is the baseline. The shading function is given by
\begin{equation}
S_i(\vecn) = \left\{
\begin{array}{cc}
0, & -\vecn\cdot\vecr_i > \sqrt{|\vecr_i|^2 - R_\moon^2}\\ 
1, & {\rm else}\\ 
\end{array}
\right..
\label{eq:shade_func}
\end{equation}

Each of the DSL daughter satellite for the low frequency interferometric observation is equipped with a pair of mutually orthogonal short tripole antennas on each side of the slab-shaped satellite \citep{chen2023}. As the antennas are electrically short, i.e. its geometric size is much shorter than the wavelength, along each of the poles direction the antenna can be treated as a short dipole, the whole system can be treated as three orthogonal short dipoles \citep{chen2018antenna}. The antennas would be well aligned--the precision of orientation control of the satellites is better than $0.1^\circ$, though as the satellites are oriented toward the center of Moon, there are some slight angle differences between the satellites, which are still much smaller than the width of the antenna beam. The slight misalignment between the antennas on different satellites could induce slight mixing between the different polarization components, but this could also be corrected as the orientation is known and the component of all three orthogonal directions are measured. However, the detailed treatment of this is beyond the scope of the present paper, so we do not consider this problem further in the present work. 

During the science observation of DSL, the satellites are lunar-center-oriented, so the orientation of each dipole changes in the inertial space as the satellite moves along the orbit. In order to form the effective visibility for a particular polarization direction in the inertial frame, one may use the combination of the differently oriented antennas. However, we find that the direction of antennas does not have a large impact on the low-resolution map-making discussed in this paper, so we simply ignore the primary beam of the tripole antenna here, instead consider the unpolarized sky temperature, and leave the discussion of the influence of antenna beam direction to future work.

The real parts and imaginary parts of thermal noise for each baseline are generated by independent Gaussian random number, with standard deviation given by
\begin{equation}\label{eq:noise}
\sigma^{\rm ns}_{ij} = \frac{T^{ij}_{\rm sys}}{\sqrt{2\Delta\nu\Delta t}},
\end{equation}
where $T^{ij}_{\rm sys}=\sqrt{ (\zeta_i^{-1} T^i_{\rm rec} + T^i_{\rm sky})(\zeta_j^{-1} T^j_{\rm rec} + T^j_{\rm sky})}$, where $T^i_{\rm sky}$ is the sky brightness temperature observed by antenna $i$,
\begin{equation}
T^i_{\rm sky} = \int T(\vecn) S_{ii}(\vecn) A_{ii}(\vecn) d\vecn.
\end{equation}
With the electrically small antenna, the impedance of the antenna is not well-matched to that of the receiver, and vary significantly with frequency. We quantify the transmitted power from antenna $i$ to its receiver by the transmission efficiency parameter $\zeta_i$, while assuming the receiver has a well defined noise temperature which stays constant or varies slowly with frequency. $T^i_{\rm rec}/\zeta_i(\nu)$  is the effective receiver noise for antenna $i$. The receiver noise is generally much smaller than the sky temperature for the frequencies considered here, even with the low transmission efficiency the sky temperature still dominates over the receiver noise for reasonable design \citep{Shi:2022xdw}. We also assume that receivers on all satellites have the same receiver noise temperature, and neglected the slight difference between $T^i_{\rm sky}$ and $T^j_{\rm sky}$ for satellites at different positions, the slight differences in these terms would not affect our result. 

The calculation of visibilities with thermal noise can be written in matrix form,
\begin{equation}\label{eq:mat_vis}
\matV = \matB\mats + \bm{\eta},
\end{equation}
where $\bm{\eta}$ is the thermal noise which we model as a Gaussian with standard deviation given by (\ref{eq:noise}), and $\matB$ is the beam matrix with its entries given by
\begin{equation}\label{eq:Bmat}
B_{\alpha\beta} = S_\alpha(\vecn_\beta) A_\alpha(\vecn_\beta) \exp\left(-2\pi i\frac{\nu}{c} \vecn_\beta\cdot \vecr_\alpha\right)\Delta\Omega_p,
\end{equation} 
where $\alpha$ denotes baselines and time, $\beta$ denotes pixels of the sky, and $\Delta\Omega_p$ is the pixel area. The $\matV$, $\mathbf{n}$ and $\matB$ are matrices of dimension $(2\times N_t\times N_{\rm baseline},)$, $(2\times N_t\times N_{\rm baseline},)$ and $(2\times N_t\times N_{\rm baseline},\, N_{\rm pix})$, respectively, where $N_t$ is the number of time points, $N_{\rm baseline}$ is the number of baselines $N_{\rm pix}$ is the number of pixels, and the factor 2 is due to storing separately the real and imaginary parts of the complex numbers.

\subsection{Finite time resolution and bandwidth}
\label{sec:finite}
The visibility data are usually averaged over a short time interval $\Delta t$ to reduce the data flow. The averaging of visibility data will cause a direction-dependent damping as given by  Eq.~(\ref{eq:vis}). 
Assuming a $\dot{\vecr}_{ij} = \vecv_{ij}$ is constant during $\Delta t$, the damping term is
\begin{eqnarray}\label{eq:damp_int}
D(\vecn) &=& {\rm sinc}\left(\pi \frac{\nu}{c} \vecn\cdot \vecv_{ij}\Delta t\right)\nonumber\\
&=& {\rm sinc}\left(\pi \frac{\vecn\cdot\Delta \vecr_{ij}}{\lambda}\right),
\end{eqnarray}
where $\Delta \vecr_{ij} = \vecv_{ij} \Delta t$ is the change of the baseline in $\Delta t$. 
Here in the simulation, we dynamically determine the sampling time for each baseline, and keep the change in baselines for two contiguous time points $|\Delta \vecr_{ij}| = \lambda/8$, where the maximum damping is ${\rm sinc}(\pi/8)\approx 0.974$ and can be ignored. We also require $\Delta t < 25\sec$ to avoid significant change in shading and response. 
The time resolution is then given by
\begin{eqnarray}\label{eq:t_resol}
    \Delta t=\min\left(\frac{\lambda}{8|\vecv_{ij}|}, 25\sec\right).
\end{eqnarray}
Nevertheless, we find that for most baselines (more than 99\%), the sampling time interval is set by the first requirement.

The actual visibility is computed over a finite bandwidth. Thus, 
\begin{eqnarray}
V^{\rm BW}_{ij}(\nu_0) &=& \frac{1}{\Delta\nu} \int_{\nu_0-\frac{\Delta\nu}{2}}^{\nu_0+\frac{\Delta\nu}{2}} d\nu ~V_{ij}(\nu)\nonumber\\
&\approx&
\int d\vecn~ T(\vecn,\nu_0) S_{ij}(\vecn) A_{ij}(\vecn)  \nonumber\\
&&\times \frac{1}{\Delta\nu} \int_{\nu_0-\frac{\Delta\nu}{2}}^{\nu_0+\frac{\Delta\nu}{2}} d\nu
\exp\left(-2\pi i\frac{\nu}{c} \vecn\cdot \vecr_{ij}\right) \nonumber\\
&=&
\int d\vecn~ T(\vecn,\nu_0) S_{ij}(\vecn) A_{ij}(\vecn) \exp\left(-2\pi i\frac{\nu_0}{c} \vecn\cdot \vecr_{ij}\right) \nonumber\\
&& \times ~
{\rm sinc}\left(\frac{\pi \vecn\cdot \vecr_{ij}\Delta\nu }{c}\right) 
\end{eqnarray}
where the second line takes the approximation that except for the phase factor, all other factors vary slowly with frequency. As shown by the last line, the signal from directions that is parallel to the baseline would be damped by the sinc function factor. For a maximum baseline of 100 km, $c/(\pi \vecn\cdot \vecr_{ij})\approx 0.95$ kHz. Thus, for greater bandwidth, there could be a de-correlation due to the finite bandwidth effect at the long baselines.

 This problem can be mitigated by computing the visibility function with  different offset delays, which is equivalent to choosing different reference points, to limit the argument of the sinc function to small values. On the other hand, if this effect is ignored and the visibility is computed only at zero offset, it will only affect the contribution to the visibility of long baselines from the sky directions close to the baseline, their influence on the overall synthesized map would still be very small, as found in \citet{Shi:2022xdw}. Furthermore, for the low-resolution map-making discussed in this paper, the information is mainly contributed by short baselines which is quite insensitive to the bandwidth smearing. We shall not consider this effect further in this work, and set $\Delta\nu=8 {\rm\ kHz}$.

\subsection{Lunar Reflection}
\label{sec:reflection}
\begin{figure}
\fig{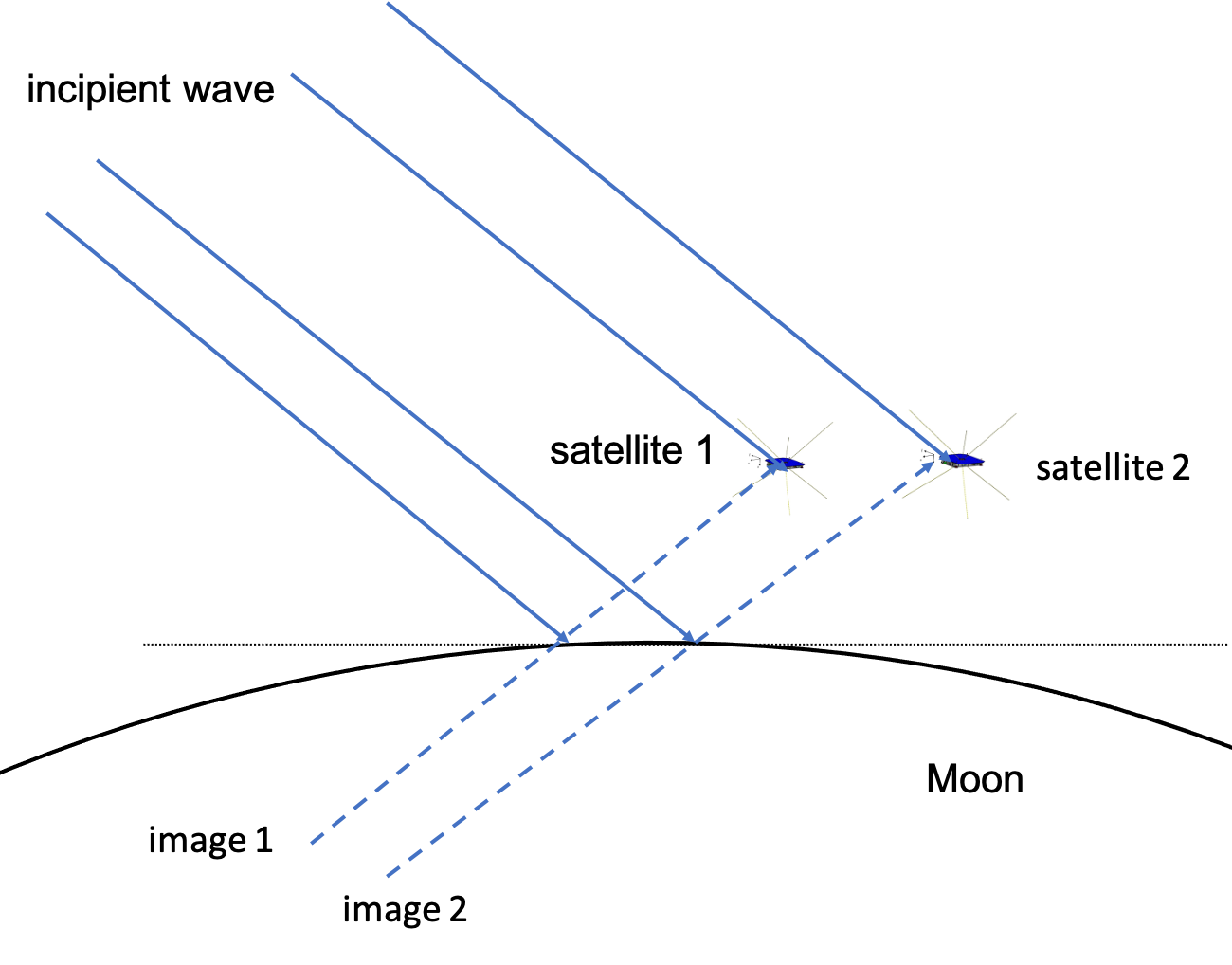}{0.45\textwidth}{}
\caption{The directions of incipient wave and reflected wave for two satellites. In the case where the height and distance between satellites are much smaller than the radius of the Moon, the images seen by the two satellites are nearly in the same direction, and the reflected wave geometric delay between the two satellites are close to that of the incipient wave.}
\label{fig:reflection}
\end{figure}

In the above we have neglected the reflection of the Moon, here we discuss the reason for this treatment. The reflection of the radio wave by the Moon is partly specular, partly diffusive, and at the long wavelength the specular component is more significant. From the perspective of a satellite, each radio source above the lunar horizon also has a corresponding reflected image. As the surface of the Moon is not completely smooth, this image may become an extended blob with multiple bright spots, or even a long bright column when the incidence angle is very low. This reflected wave is however delayed by at least $2h/c$ with respect to the directly incipient wave, where $h = 300$ km is the height of the satellite and $c$ the speed of light. This is much longer than the delay for even the longest baseline of the array (100 km). So, even with the delayed correlation computation outlined above in Section ~\ref{sec:finite}, coherence will be lost between the incipient wave and the reflected wave, the reflected wave will behave as a separate source from the direction of the image. As the satellites move around the Moon, however, the positions of these image will change. But note that for such an image, the geometric phase difference induced on a pair of satellite would be approximately the same as the original incident wave. This image would be at the mirror image position if the surface of the Moon is a flat plane, and thus produce the same geometric phase difference as the original source, but as the Moon is spherical, this is only an approximation, valid for satellites flying at the same height, and in the limit that both the heights and the distances between the satellites are small compared with the radius of the Moon. So in the real case the image position would deviate from the position for flat mirror, and change with time, as depicted in Figure \ref{fig:reflection}.  

In our reconstruction program, we apply the shading function of Eq.~(\ref{eq:shade_func}), which assumes that there is no radiation from the direction of the Moon, so this contribution to the visibility from reflected wave will be attributed to the original incipient wave. It has been estimated that the power of the reflection is $\xi_R \approx 7\%$  of the original times the original \citep{Evans_1969}, then in a simple treatment we may ignore the slight direction difference for the two satellites, and estimate that the received power from the source will be $1+\xi_R$ times the original. Then we can reconstruct image without considering the reflection, and only correct it at the end by dividing $1 + \xi_R$.

The actual reflection can be more complicated. As we noted, there are differences in the direction of the reflected wave for the different satellites, and furthermore the reflection coefficient may depend on the incipient angle, and the undulation of the lunar surface generate non-uniform reflections. These are beyond the scope of the present work, and will be investigated in subsequent works, but note that as discussed above, the reflected wave is not coherent with the incipient wave, and its intensity is at least an order of magnitude smaller than the incipient wave. Its main effect is slightly increase the flux of sources in the reconstructed image and also a slightly increase of the noise. In this paper we will ignore these effects.

\section{Map Reconstruction and the Aliasing Effect}
\label{sec:map}

\subsection{Reconstruction Algorithm}

The full-sky map $s(\vecn)$ can be reconstructed by solving the linear equation (\ref{eq:mat_vis}) \citep{Huang2018,Shi:2022xdw}. However, due to the incomplete $uvw$ coverage or noise, this problem is often ill-posed, i.e. without a unique solution, and a pseudo-inverse or regularization procedure is often necessary. In this paper, we solve Eq.~(\ref{eq:mat_vis}) using the Tikhonov regularization  \citep{2017MNRAS.465.2901Z,Yu:2023idm}, which is obtained by minimizing 
\begin{equation}\label{eq:bay_mapmk}
L=(\matB\matsnew - \matV)^T \matN^{-1} (\matB\matsnew - \matV) + (\matsnew-\mats_p)^T\matR(\matsnew-\mats_p),
\end{equation}
where $\matN = \langle \bm{\eta^\dagger \eta}\rangle $ is the covariance matrix of the noise, which is assumed to be a diagonal matrix in our simulation, $\mats_p$ is a prior sky map, and $\matR$ is the regularization matrix. Minimization of the first term 
seeks the solution which would yield the observed data, while the second term penalizes deviation from the prior map.
The solution is given by
\begin{equation}\label{eq:mapmk}
\matsnew = (\BTB + \matR)^{-1} \BT(\matV - \matB \mats_p) + \mats_p,
\end{equation}
where $\BTB$ is the direction-dependent dirty beam, the addition of $\matR$ makes it less singular and thus can be inverted at the price of a slight bias, and $\BTV$ gives the so-called dirty map, so this inversion can also be viewed as an attempt to deconvolve the dirty map.

Below we consider the case with no prior, i.e. $\mats_p=0$, and a uniform $\matR$ matrix that is proportional to the identity matrix, $\matR =\varepsilon \matI$. 
In this case, we can decompose the dirty beam $\BTB$ by its eigenvectors, 
\begin{equation}\label{eq:eigen_decomp}
\BTB = \matQ \matW \matQ^T,
\end{equation}
where $\mathbf{W}$ is a diagonal matrix containing the eigenvalues of $\BTB$, while $\mathbf{Q}$ are the matrix made up of corresponding eigenvectors. The larger the noise, the smaller the diagonal elements of $\matW$. Because $QQ^T=\matI$, we have $\BTB + \matR = \matQ (\matW + \varepsilon\matI)\matQ^T$. Therefore, the inversion of $\BTB+\matR$ can then be written as 
\begin{eqnarray}\label{eq:eigen_decomp_reg}
(\BTB + \matR)^{-1} &=& \matQ (\matW + \varepsilon\matI)^{-1} \matQ^T.
\end{eqnarray}
Without the regularization matrix, those modes with $W_{ii} < 1$ would 
be amplified and create an instability. The regularization $\matR =\varepsilon \matI$ avoids this instability. We could link this to the largest eigenvalue/singular value of the matrix, 
$\varepsilon=\epsilon W_{\rm max}$, where $W_{\rm max}$ is the maximum eigenvalue of $\BTB$,  $\epsilon$ is the regularization parameter.

We first ignore the thermal noise and consider the reconstruction error introduced by regularization.  Substituting Eq.~(\ref{eq:eigen_decomp}), Eq.~(\ref{eq:eigen_decomp_reg}) and $\matV=\matB\mats$ into Eq.~(\ref{eq:mapmk}) with $\mats_p=0$, we find $\matsnew=\matB_{\rm eff} \mats$, where
\begin{eqnarray}\label{eq:beam_eff}
\matB_{\rm eff} &\equiv& (\BTB + \matR)^{-1} \BTB\nonumber\\
&=& \matQ(\matW + \epsilon W_{\rm max}\matI)^{-1}\matW\matQ^T,
\end{eqnarray}
is the `effective beam' that is convolved with the input sky map to produce the reconstructed one. 
In the case of perfect inversion $\matB_{\rm eff}=\matI$, while $(\matB_{\rm eff}-\matI)$ indicates the imperfection of the inversion. 

To ensure numerical stability, a sufficiently large $\epsilon$ must be chosen. However, the deviation of $\matB_{\rm eff}$ from $\matI$ would be greater for larger $\epsilon$. A number of methods, such as the  generalized cross validation\citep{Yu:2023idm,golub1979} and L-curve \citep{Yu:2023idm,eda2022,eastwood2018}, have be proposed to choose an optimal $\epsilon$ value.

\begin{figure*}
\centering
\fig{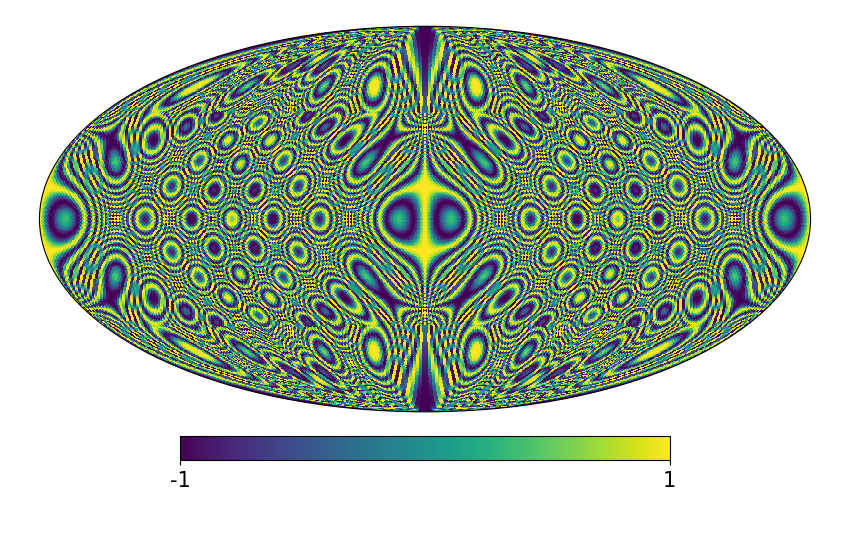}{0.45\textwidth}{(a) The pixel-center method }
\fig{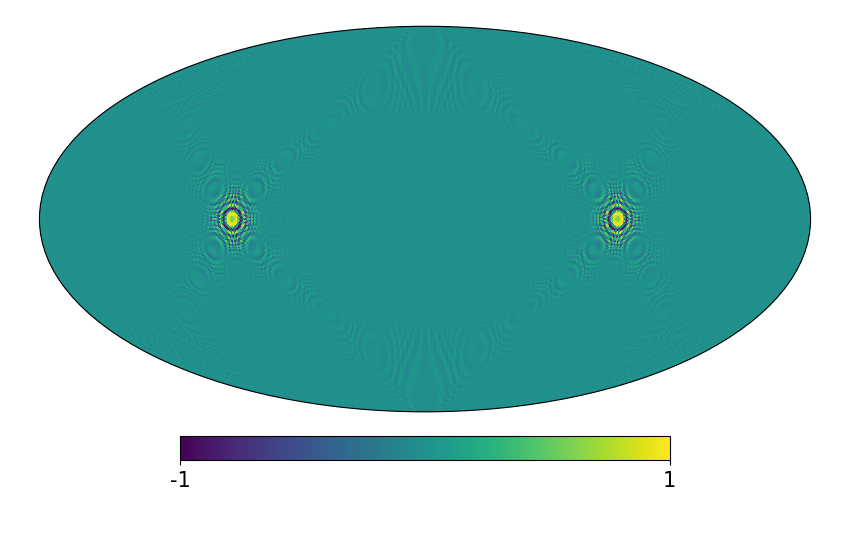}{0.45\textwidth}{(b) The pixel-averaging method.}
\caption{The comparison of a row of the $\matB$ matrix, computed with the pixel-center method (a) and that with the pixel-averaging method (b). 
In each panel, the real part of fringe term $\exp\left(-2\pi i\frac{\nu}{c} \vecn_\beta\cdot \vecr_\alpha\right)$ is shown with $\nside=64$. The baseline is assumed to be aligned with $y$-axis, with a length of  $\approx 250.2\lambda$, corresponding to $b\approx 4 b_p$. The shading of Moon is not included in this plot. 
}
\label{fig:av_vs_ct_beam}
\end{figure*}

\subsection{The Aliasing effect}
For a sky with a total pixel number of $N_\pix$, the size of the dirty beam matrix $\BTB$ scales as $N_\pix^2$, and the computational complexity of the matrix inversion scales as $N_\pix^3$, so the pixel number is necessarily limited by the available computing resource. However, a baseline of length $b$ is sensitive to an angular size of $\lambda/b$ (in radians), where $\lambda$ is the wavelength, the corresponding sky area is $\Delta\Omega \sim (\lambda/b)^2$. Conversely, for a pixel with sky area $\Delta\Omega_p \sim \theta_p^2$ where $\theta_p$ is the characteristic side length of the pixel, the corresponding baseline length  is given by  $b_p\sim \lambda/\theta_p$. 
For example, at 10 MHz, in the case of HEALPix $\nside=64$, the baseline length corresponding to the pixel size is $b_p =\lambda/\theta_p\approx 1.87\km$. The baselines longer than this will be sensitive to the sub-pixel variations, and in the simulation and map-making process such variations can have an aliasing effect, generating artifacts on the final map, and increase the error of map reconstruction. 

\begin{figure*}
\fig{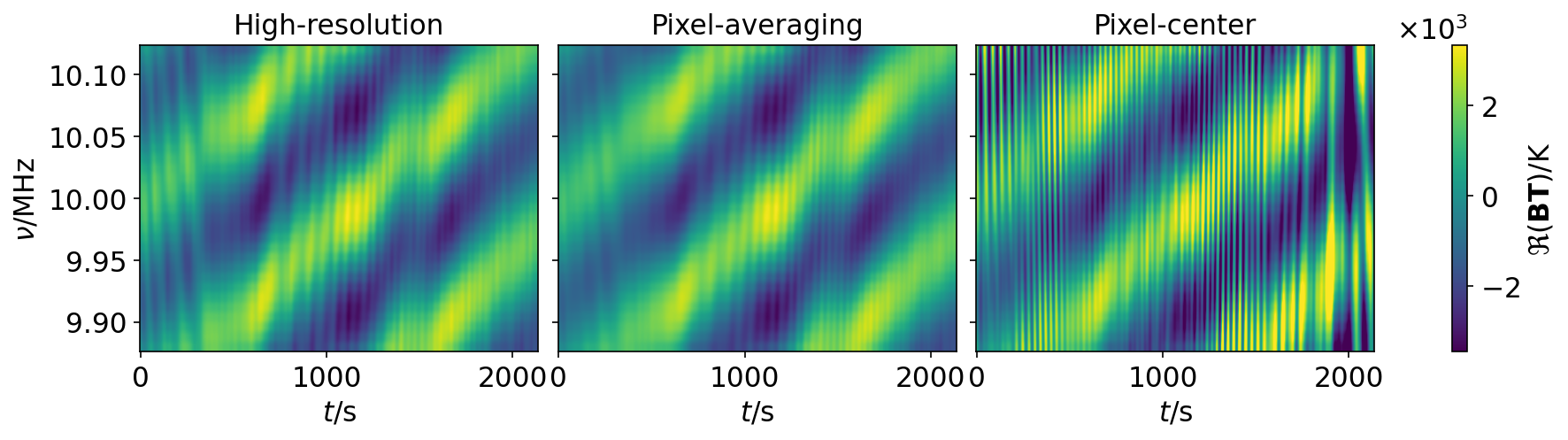}{\textwidth}{}
\caption{Comparison of the real part of the mock visibility $\Re\left(\matB\mats\right)$ computed with a high resolution of $\nside=1024$ (left panel), with that evaluated at a low resolution of $\nside=64$ using the pixel-averaging method (middle panel) and the pixel-center method (right panel) method, respectively. The baseline length is $b\approx b_p$ for $\nside=64$.
}\label{fig:vis_av_vs_ct}
\end{figure*}

This aliasing effect is manifestly visible in the construction of the beam matrix $\matB$ needed for the sky map reconstruction. In Figure \ref{fig:av_vs_ct_beam}, we plot a row of the $\matB$ matrix for a long baseline, i.e. the coefficients of sky pixels for a given visibility. Each coefficient corresponds to a pixel, so we can plot them as a map. For the purpose of visualization, we only show the fringe term $\exp\left(-2\pi i\frac{\nu}{c} \vecn_\beta\cdot \vecr_\alpha\right)$ which is oscillating and produces small-scale power, and ignore the shading of the Moon. 
A {\it naive} way to construct the $\matB$ matrix is to evaluate Eq.~(\ref{eq:Bmat}) for each matrix element, at the center of each pixel, below we shall call this the {\it pixel-center method}. However, as shown for example in Figure \ref{fig:av_vs_ct_beam} (a) for a map with \texttt{HEALPix} $\nside=64$, this results in some strange-looking patterns. These are the aliased sub-pixel fringes of that baseline, which are produced when the spatial frequency of the fringes is higher than the Nyquist frequency of the pixels. 

To overcome this problem, we can generate the beam maxtrix $\matB$ by first evaluate Eq.~(\ref{eq:Bmat}) in pixels of a higher resolution scheme, so that the angular scale probed by the given baseline is resolved by the pixels, and then down-grade to the target resolution by averaging neighboring pixels. We shall call this the {\it pixel-averaging method}. 
For example we can start with  $\nside=1024$, and then downgrade to $\nside=64$ by averaging, then we can obtain a normal-looking result, as shown in Figure \ref{fig:av_vs_ct_beam} (b), where the response on scales smaller than the pixel size has been smoothed out, and a normal-looking figure is produced. Rings of fringe is only seen near the two ends of the baseline direction on the spherical sky, because the phase variation of $\matB$ are slow near these two directions. 

Of course, if one adopts a very fine pixelization such that all baselines are only sensitive to scales above the pixel size, the aliasing effect would not be important. But this would produce huge matrices. For $\nside=1024$, the full $\BTB$ requires $\sim 633$ terabytes to store for single precision floating number, while for $\nside=64$ the matrix size is reduced to $\sim 9.7$ gigabytes, and make the consequent imaging process affordable.

\begin{figure*}
\centering
\gridline{
\fig{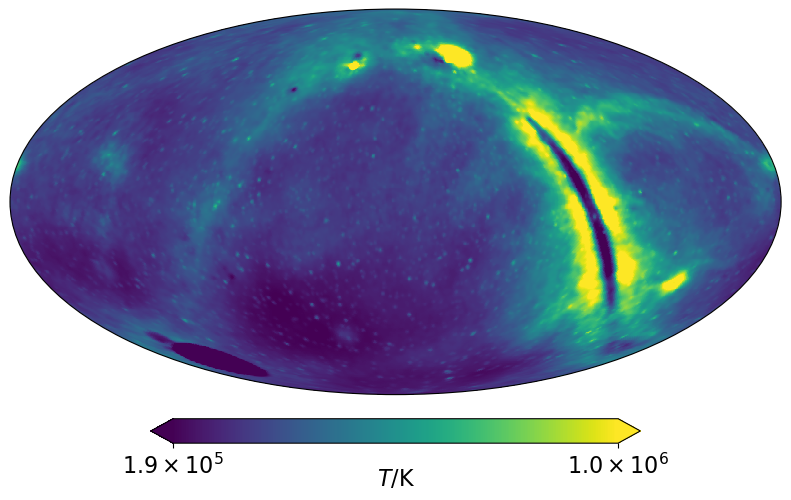}{0.4\textwidth}{(a) Input map}}
\gridline{
\fig{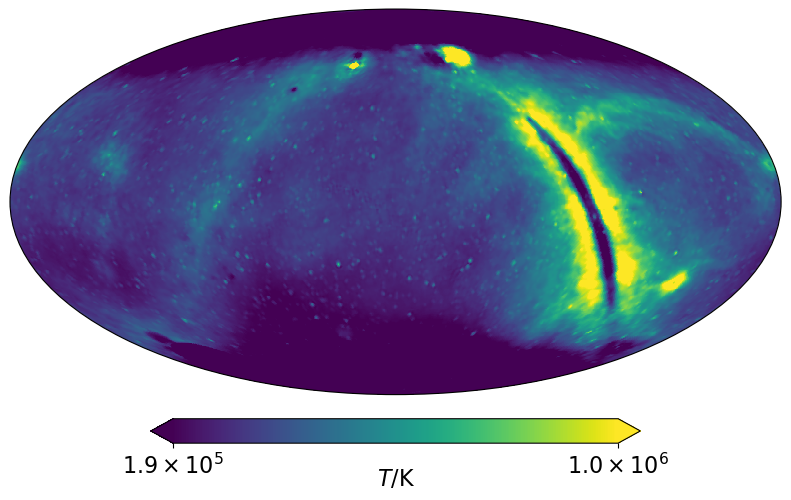}{0.4\textwidth}{(b) Map reconstructed without noise}
\fig{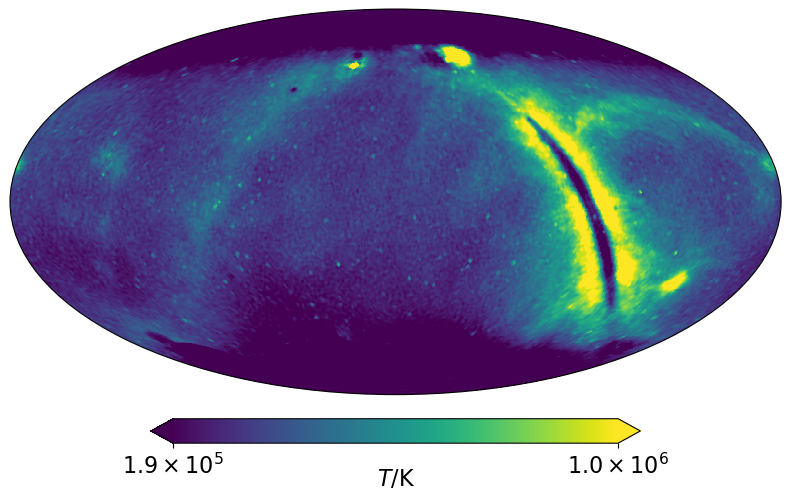}{0.4\textwidth}{(c) Map reconstructed with noise}
}
\gridline{
\fig{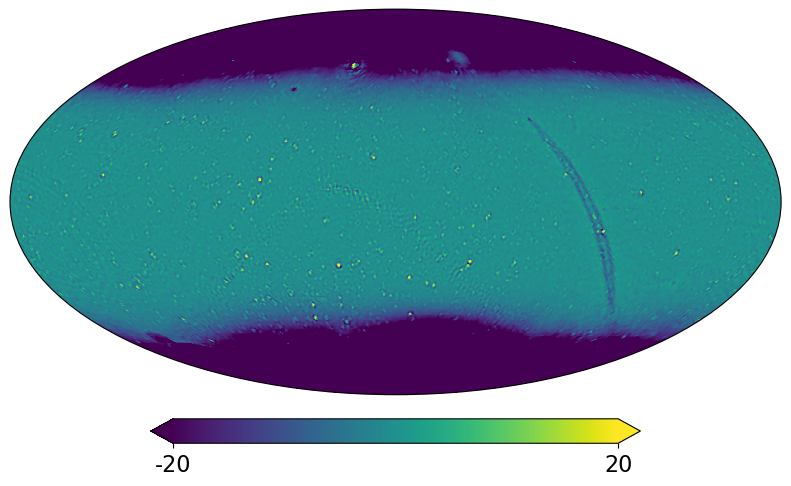}{0.4\textwidth}{(d) Relative percentage error without noise}
\fig{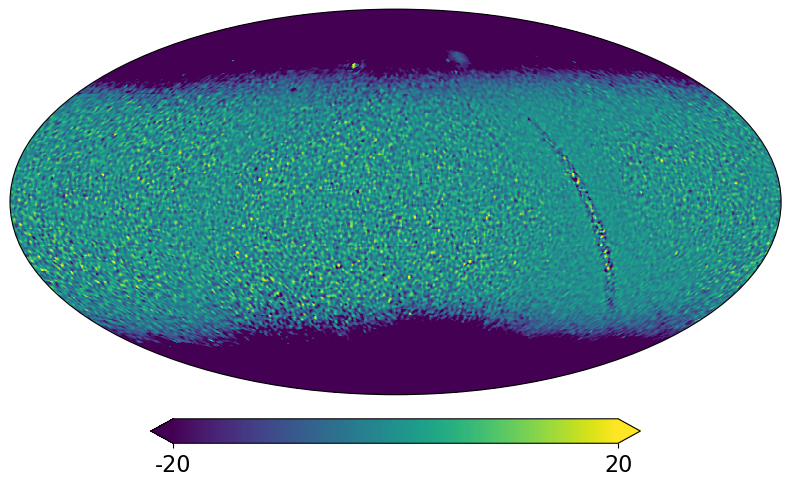}{0.4\textwidth}{(e) Relative percentage error with noise}
}
\caption{The input and reconstructed maps at 10 MHz, with the regularization parameter $\epsilon=10^{-4}$ and baseline combination $b<2b_p$.
The upper panel shows the input sky map. The middle-left panel shows the reconstructed map without thermal noise, and the middle-right panel shows the reconstructed map with thermal noise. The lower-left panel shows the percentage relative error between the input and reconstructed map without thermal noise, and the lower-right panel shows the relative error with thermal noise. }\label{fig:skymap_re_10MHz}
\end{figure*}

This aliasing also affects the the mock visibilities we generate in the simulation. The real part of a mock visibility for a baseline length $b_p \approx \lambda/\theta_p$ is shown by the waterfall plot in Figure \ref{fig:vis_av_vs_ct}. The left panel shows the visibility $\matB \mats$, computed using the beam matrix $\matB$ and mock sky map at the high resolution $\nside=1024$. The right panel shows the same visibility, but produced by the pixel-center method at the resolution of $\nside=64$. The middle panel shows the same visibility obtained via the pixel averaging method, also with final resolution of $\nside=64$. Compared with the visibility generated in the high resolution case (left panel), the mock visibility by the pixel-center method show some strong deviations, which are artifacts of the limited pixel resolution, while the pixel-averaging method reproduce the large scale variations well, though the small scale variations are smoothed out.
Of course, this aliasing effect on visibility data does not exist in the real data, but for map reconstruction the aliasing effect in the $\matB$ matrix still needs to be taken into account for the real data processing. 

The pixel-averaging method described above is equivalent to convolving the high-resolution map with a pixel-shape window and then re-sample it at the center of the low-resolution pixels. This convolution damps the sub-pixel variations while conserve the large-scale response in the $\matB$ matrix. Although the pixel-shape of the \texttt{HEALPix} is direction-dependent, hence the pixel-averaging is only an approximation to an exact convolution, it is computationally simpler and sufficiently accurate, so we adopt this method.

In our simulation we generate visibilities with $\nside=1024$, which has sufficient resolution to ensure that the mock visibility is close to the real data, then reconstruct the map from the visibility data. To suppress the discontinuity caused by the pixelization in the reconstructed map, we first make a map with $\nside=256$ (resolution $\sim 13.7\arcmin$) by splitting one pixel into 16 sub-pixels, then convolve it with a Guassian window with Full-Width-Half-Maximum (FWHM) of 1 degree in the image space to produce the final map.

Below we evaluate how the reconstructed map is affected by the aliasing effect under the expected thermal noise and the sub-pixel power,  comparing the results of the pixel-averaging method with the pixel-center method.

\begin{figure*}
\centering
\gridline{
\fig{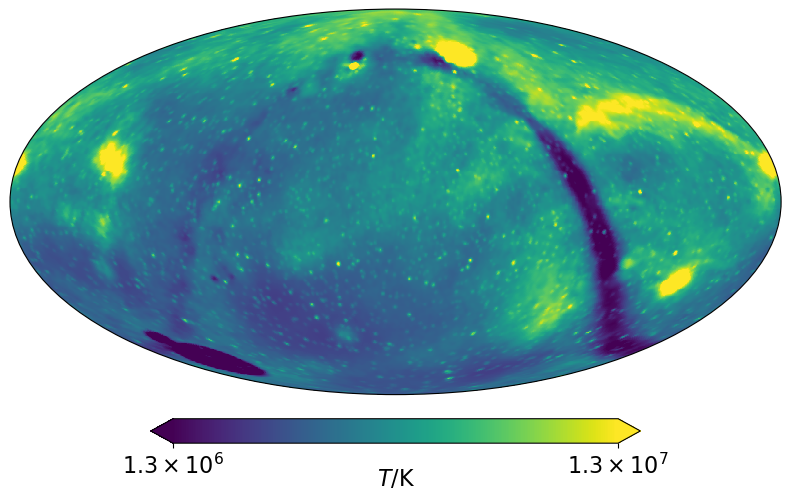}{0.4\textwidth}{(a) Input map}}
\gridline{
\fig{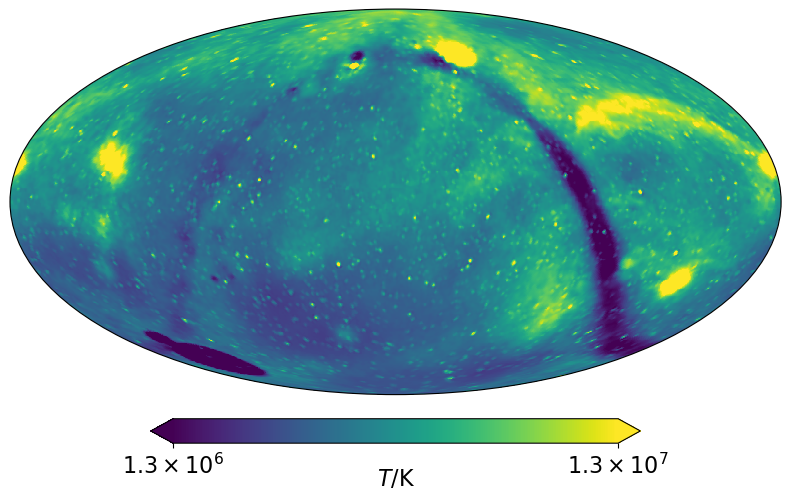}{0.4\textwidth}{(b) Map reconstructed without noise}
\fig{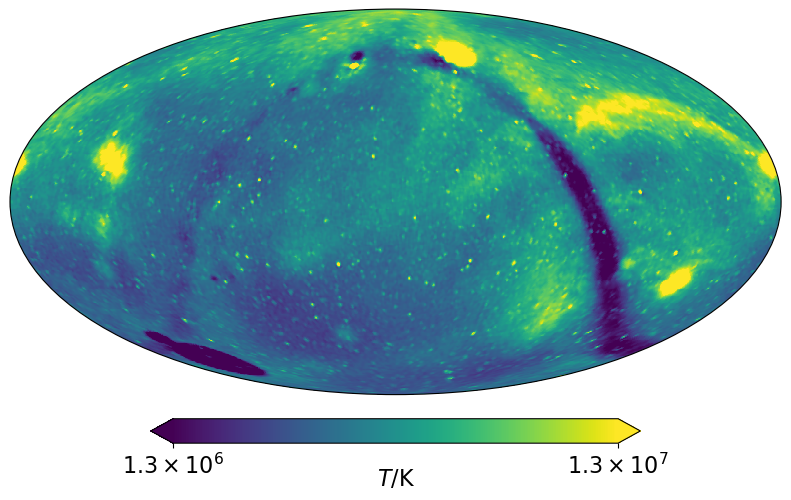}{0.4\textwidth}{(c) Map reconstructed with noise}
}
\gridline{
\fig{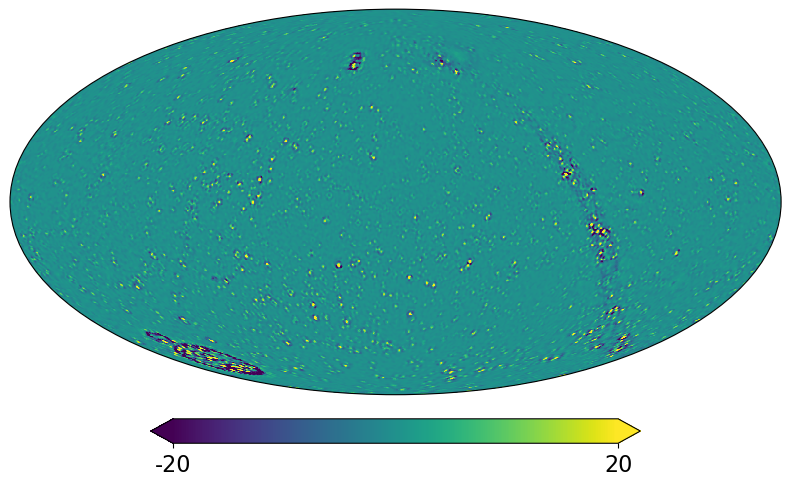}{0.4\textwidth}{(d) Relative percentage error without noise}
\fig{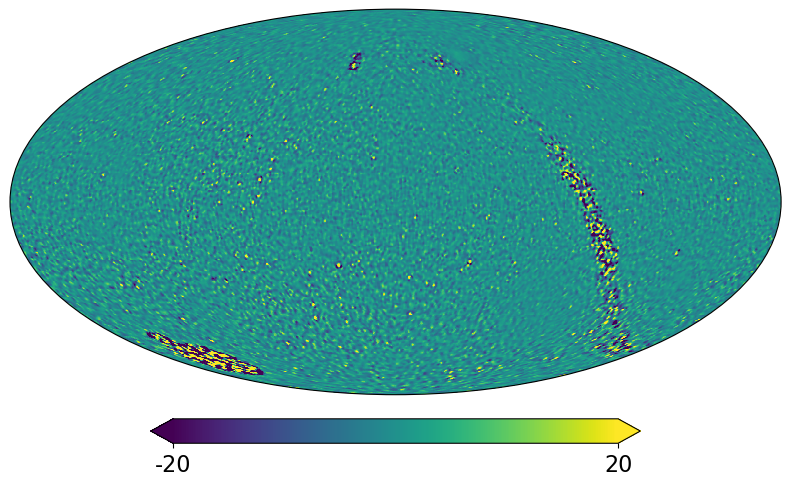}{0.4\textwidth}{(d) Relative percentage error with noise}
}
\caption{Same as Figure \ref{fig:skymap_re_10MHz} but at 3 MHz, with baseline combination $b<2b_p$ and $\epsilon=10^{-6}$. 
}
\label{fig:skymap_re_3MHz}
\end{figure*}

\subsection{Reconstruction Result on Global Scale} 
\label{sec:global}
Assuming one full precession period (1.3 year), we make a simulated sky map reconstruction at 10 MHz. We adopt an $\epsilon=10^{-4}$, and use visibilities for all baselines with $b<2b_p$. In Figure \ref{fig:skymap_re_10MHz}, we plot (a) the input map; (b) the map reconstructed without noise; (c) the map reconstructed with noise; (d) the relative error for the case without noise, i.e. the map in (b); and (e) the relative error for the case with noise, i.e. the map in (c).

\begin{figure*}
\centering
\gridline{
\fig{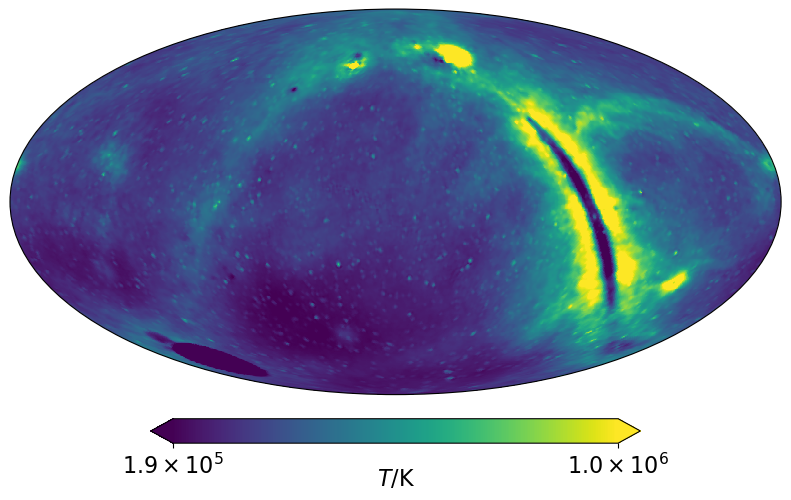}{0.4\textwidth}{(a) Map reconstructed without noise}
\fig{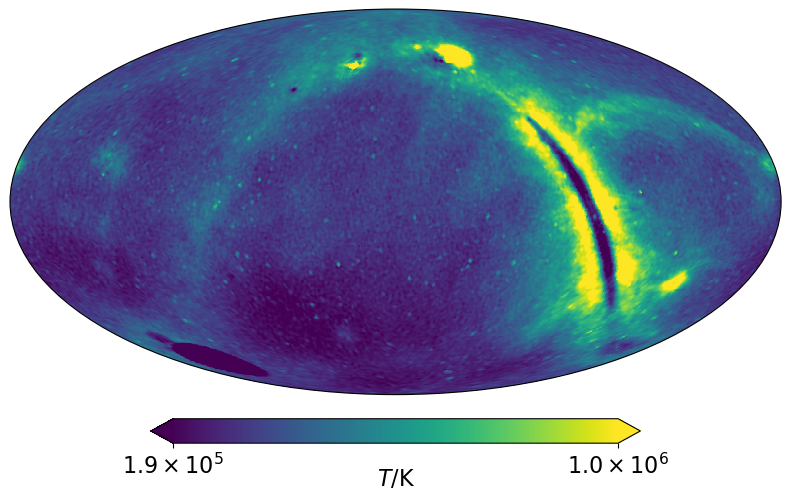}{0.4\textwidth}{(b) Map reconstructed with noise}
}
\gridline{
\fig{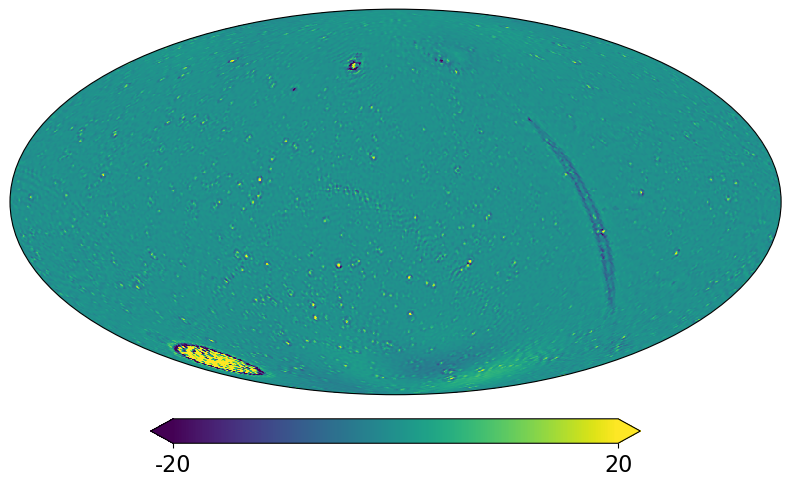}{0.4\textwidth}{(c) Relative percentage error without noise}
\fig{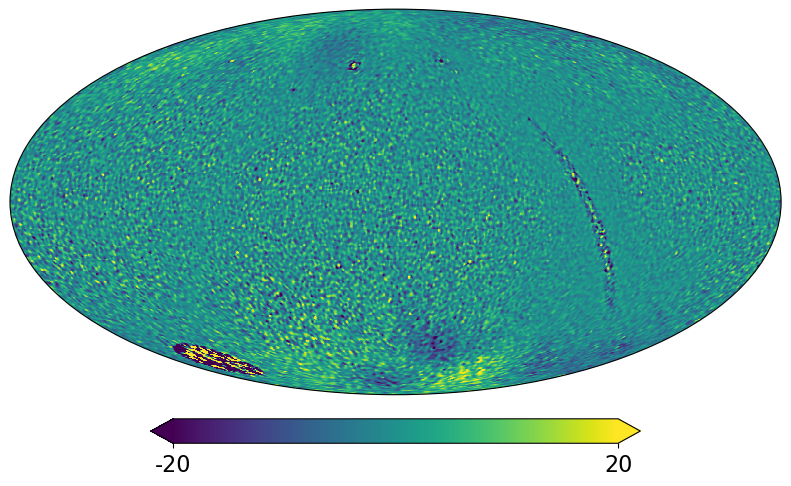}{0.4\textwidth}{(c) Relative percentage error with noise}
}
\caption{The 10 MHz maps reconstructed with a non-zero prior map (upper panels), and the fractional errors (lower panels). The input map is the same as before. In each row, the left panel shows the result without thermal noise, and the right panel shows the result with thermal noise.
}\label{fig:skymap_re_10MHz_prior}
\end{figure*}

A most striking feature of the reconstructed maps is that the polar regions are much darker in the reconstructed map, i.e. there is a systematically lower average temperature in the polar regions. In the relative error map, there is a negative bias error. This is due to the lack of short projected baselines at the polar region. Since the baselines are oriented along the tangents of the orbit, the inclination angle is $30^\circ$, so towards the direction of the poles, the shortest projected length of the baseline is $\cos 30^\circ =\sqrt{3}/2$ of the baseline length. In contrast, the reconstruction error at equatorial region in Figure \ref{fig:skymap_re_10MHz} is small, and the relative error is within $20\%$ because the shortest projected baseline length can reach 0 at the equatorial region. However, for input maps without small scale power,  one can take smaller $\epsilon$ values, and the map can be better reconstructed without this problem appearing \citep{chen2023}.

Despite this bias, at most places the map is well reconstructed, with small errors. Near the bright galactic plane, there is some more apparently larger error. In the case with noise, we can also see more randomly distributed errors in the map, but they are uniformly distributed over the sky.

\begin{figure*}
\centering
\gridline{
\fig{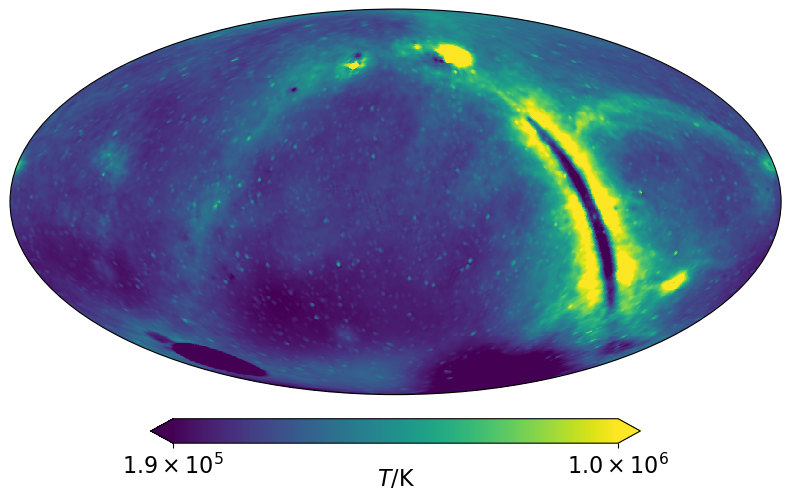}{0.4\textwidth}{(a) Map reconstructed without noise}
\fig{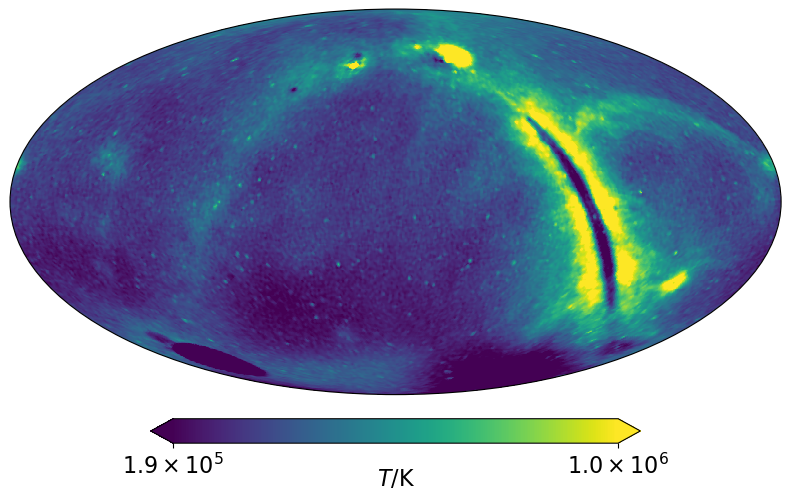}{0.4\textwidth}{(b) Map reconstructed with noise}
}
\gridline{
\fig{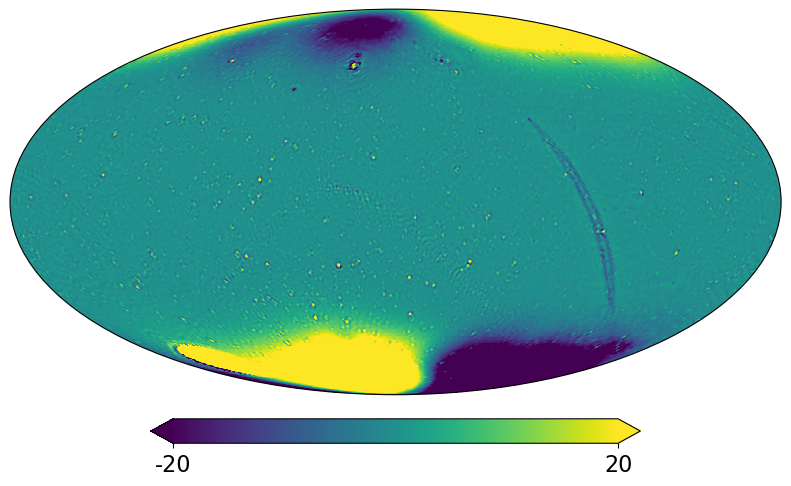}{0.4\textwidth}{(c) Relative percentage error without noise}
\fig{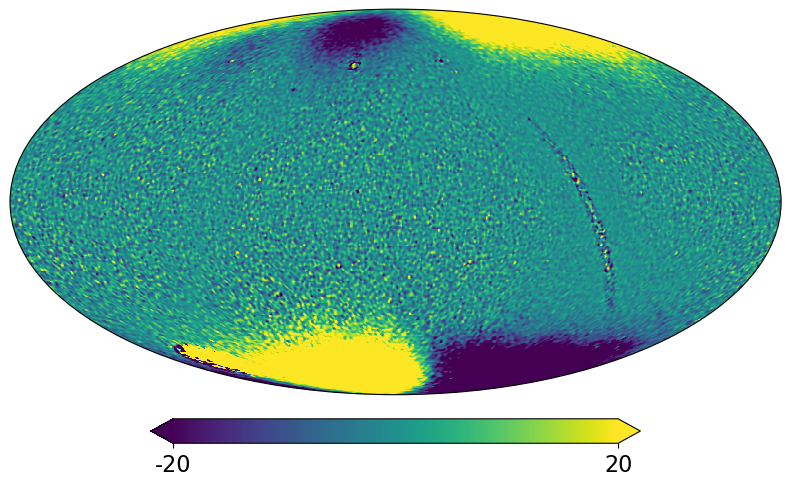}{0.4\textwidth}{(c) Relative percentage error with noise}
}
\caption{The reconstructed maps (upper panels), and the fractional error maps (lower panels) using a non-zero but incorrect prior map at 10 MHz. The input map is the same as before. 
In each row, the left panel shows the result without thermal noise, and the right panel shows the result with thermal noise.
}\label{fig:skymap_re_10MHz_prior_incorrect}
\end{figure*}

At lower frequencies, $r_{\rm proj,min}/\lambda$ will be smaller and the loss of large-scale structures will be less severe. For example, at 3 MHz, $r_{\rm proj,min}\sim 0.87\lambda$. The reconstructed all-sky maps and the corresponding fractional errors are shown in Figure \ref{fig:skymap_re_3MHz}, with all baselines $b<2b_p$ used in the construction. Note that due to the longer wavelength, the number of baselines here are also more than the corresponding case of 10 MHz, and a smaller regularization parameter can be used. Here we adopt $\epsilon=10^{-6}$. In this case the polar region is no longer `darkened'. The thermal noise in the reconstructed image is nearly uniformly distributed in most of the sky.

Another way of avoiding this darkened polar region is by employing a prior map.
As indicated by Eq.~(\ref{eq:mapmk}), a prior map could be introduced in the reconstruction. In the above reconstruction, the prior map is set to zeros, and the lack of short projected baselines for the polar regions leads to a lacking of large-scale power there. If we introduce a prior map, it could alleviate the power loss on large scales. To demonstrate this idea, we smooth the input map with a Gaussian filter FWHM$=5^\circ$ and use it as the prior map, and reconstruct the sky with the same parameter settings. The resulting reconstructed map and the percentage relative error are shown in Figure \ref{fig:skymap_re_10MHz_prior}. The incorporation of prior significantly reduced the power loss in the polar regions while the equatorial regions are still well reconstructed. However, we could see some correlated structures near the poles in the relative percentage error maps, because although the prior map help to reduce the overall negative bias, the fact remains that there are less short projected baselines for the polar region  to provide observational constraint. 

The prior map used above is a smoothed version of the input map, which is fundamentally ``correct'', but in the real world our input map may have errors. To study its impact on the reconstruction,  we also generated an incorrect prior map, by injecting Gaussian random numbers into the smoothed version of the input map in spherical harmonic space, while keeping its angular power spectrum unchanged. 
In this exercise, we keep the correlation coefficient between the incorrect prior map and the input map independent of $l$ by setting the amplitudes of random numbers at different $l$ modes. 

The reconstructed results using a prior map with a correlation coefficient of 0.8 are shown in Figure \ref{fig:skymap_re_10MHz_prior_incorrect}. 
The equatorial region is again well reconstructed, as we have sufficient information from the visibility data to update these regions. In polar regions, the loss of large-scale power is alleviated as shown by the reconstructed map, but the relative error is greater than 20\% due to the incorrect prior map. Note that the pattern of the reconstruction error is sensitive to the specific realization of the prior map, since the large-scale information in these regions is mainly contributed by the prior map.

At the higher frequencies, 
The lack of short baseline is more severe, but these are also bands  closer to the frequency band accessible to ground-based observations. Therefore, good prior maps can be obtained by extrapolating high-frequency maps measured by ground-based telescopes. At lower frequencies, the interstellar absorption becomes stronger, and deviation from the extrapolation becomes larger \citep{2021ApJ...914..128C}, but as the projected baselines become effectively `shorter' in wavelength units, this effect is also less severe, as shown in Figure \ref{fig:skymap_re_3MHz}. 
Furthermore, the incorporation of auto-correlations which measure the large-scale power of the sky could also help to alleviate the power-loss on large scales. 

\subsection{Reconstruction Result on small patches}

\begin{figure}
\fig{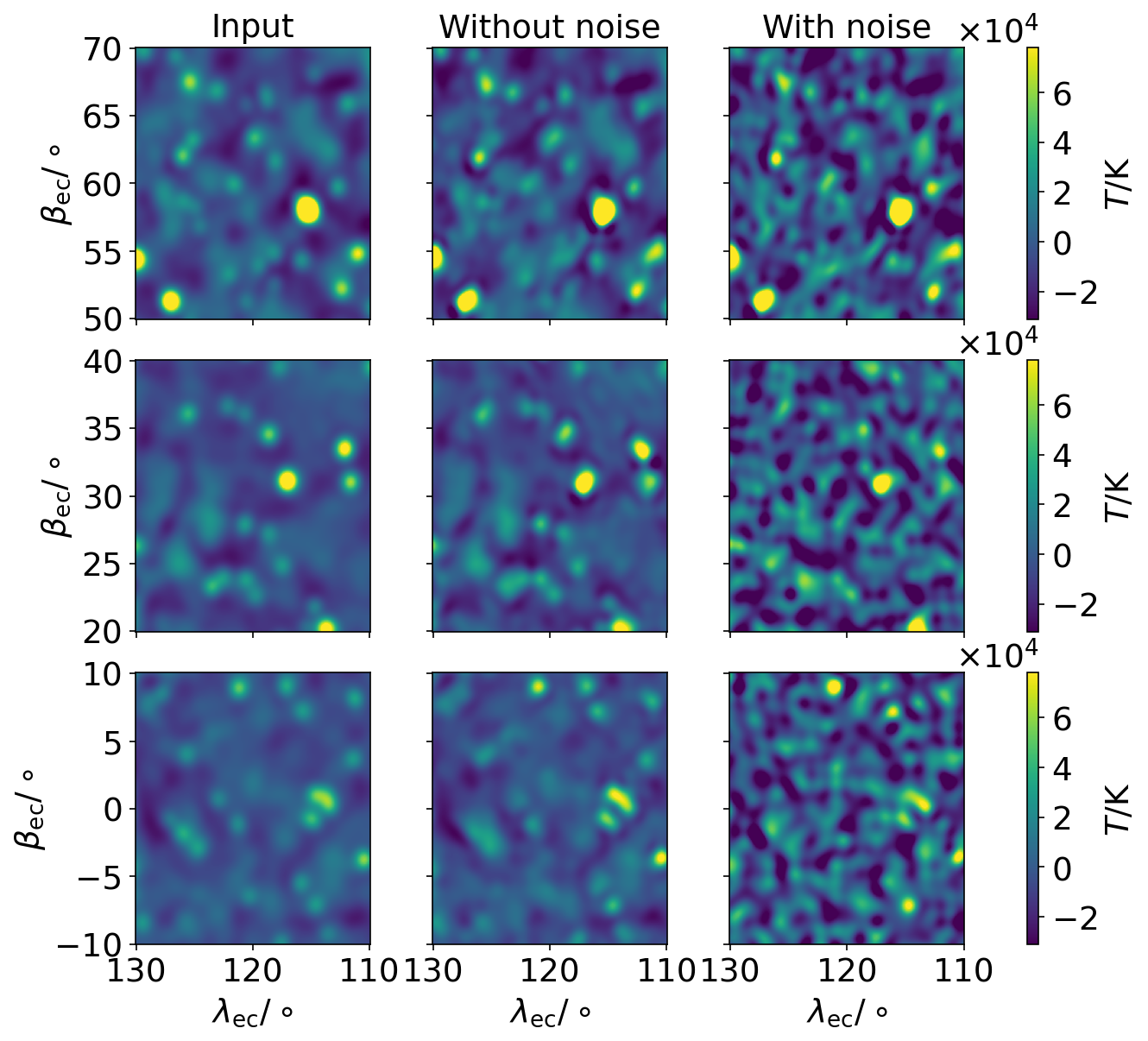}{0.48\textwidth}{}
\caption{Three $20\deg\times 20\deg$ sky patches of the input and reconstructed sky maps at 10 MHz. The input map, reconstructed maps without thermal noise and with thermal noise are shown in left, middle and right panels, respectively. 
}.\label{fig:sky_patch_10MHz}
\end{figure}

The input and reconstructed maps at 10 MHz for three $20\deg\times 20\deg$ sky patches are shown in Figure \ref{fig:sky_patch_10MHz}. The three patches are outside of the Galactic plane, and centered at longitude $\lambda_{\rm ec}=120^\circ$, latitudes $\beta_{\rm ec}=0^\circ,\ 30^\circ$ and $60^\circ$, and marked by the red boxes in Figure \ref{fig:full_sky_map}. As noted earlier, in this simulation the input maps have small scale power, but for visual comparison, to avoid the influence of diffuse structures on larger scales, the input map displayed here is smoothed to match the resolution of the reconstructed map, and the large scale diffuse components are subtracted. This is done by subtracting from the original map a smoothed map. The effective resolution of the map can be estimated from the cross angular power spectrum between the input and reconstructed map, $w_l= C_l^{\rm X}/C_l^{\rm in}$, and then we transform it to real space, which yields an FWHM$\approx 1.3^\circ$, so to remove the diffuse component we smooth the input map with a Gaussian filter of larger scale, in this case FWHM$=5^\circ$ is used.

\begin{figure}
\fig{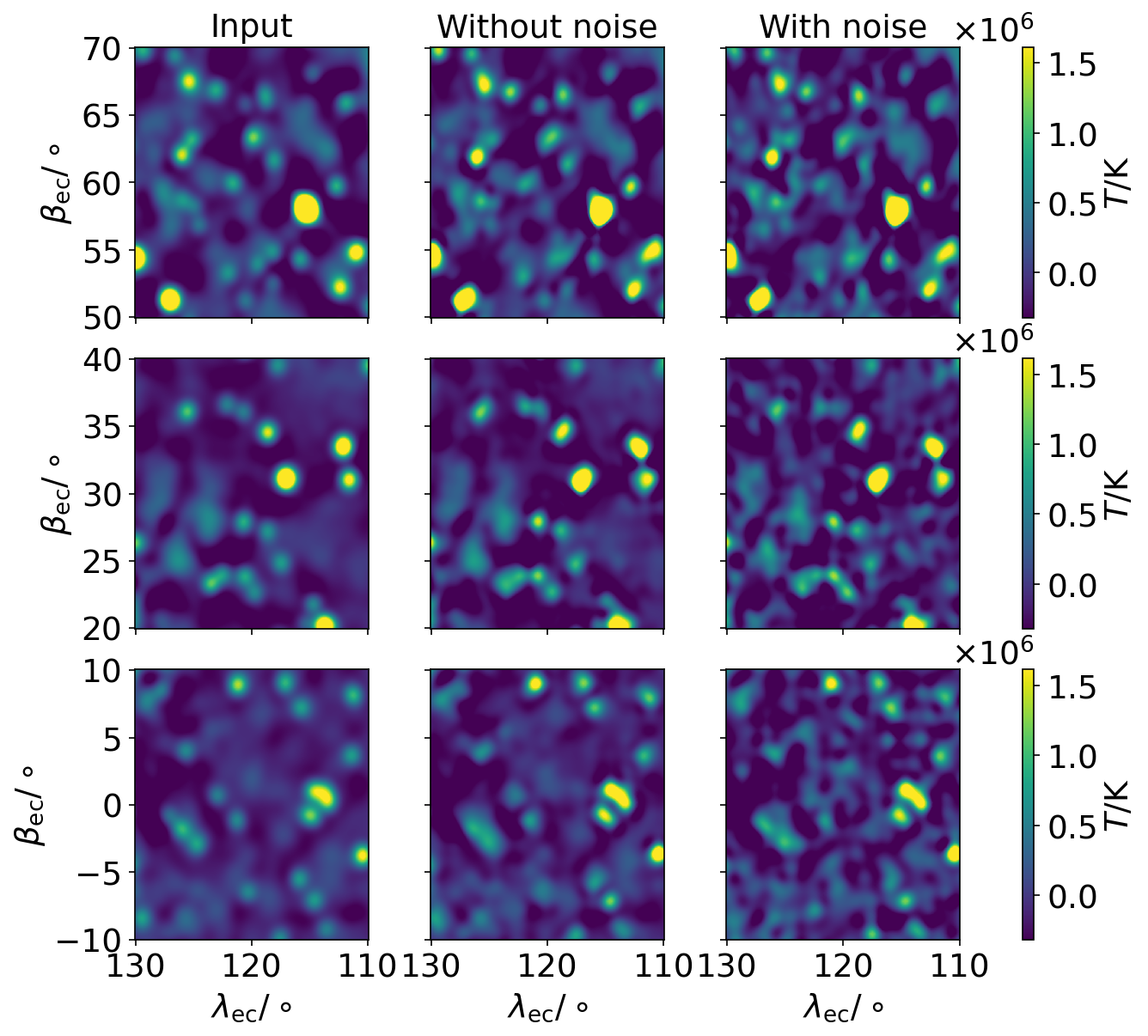}{0.48\textwidth}{}
\caption{Same as Figure \ref{fig:sky_patch_10MHz} but at 3 MHz, with baseline combination $b<2b_p$ and $\epsilon=10^{-6}$.
}.\label{fig:sky_patch_3MHz}
\end{figure}

As can be seen from the figure, without thermal noise, the small-scale structures are well reconstructed, which may even appear a little sharper than the original, which is a result of the Gaussian filtering.  Some point sources are stretched along one direction, which is due to the finite pixel size $\sim 1^\circ$ of the reconstructed map, though this can be suppressed in post-processing. However,  the image quality is somewhat degraded when thermal noise is added. The brighter sources are still visible, but then there are also noise structures which appear, and  the fainter structures are swamped by these noises in the image. This effect limits the sensitivity of the sky image made from the survey. 

A somewhat surprising result emerges at lower frequency. In Figure \ref{fig:sky_patch_3MHz} we show the reconstructed maps at 3 MHz. We might think that at the lower frequencies the system temperature is higher and so the image would be noisier, but in Figure \ref{fig:sky_patch_3MHz} the images appear to be better reconstructed than those in Figure \ref{fig:sky_patch_10MHz} with the presence of noise. A possible reason for this is, although at lower frequency, the system temperature which is mainly contributed by the diffuse component is higher, the brightness temperatures of the strong sources are also higher. Moreover, because the strong absorption at lower frequency near the Galactic plane, the increment of the system temperature is actually not as much as the strong sources outside the Galactic plane, so for these patches, the  bright sources have greater signal-to-noise ratios at 3 MHz than at 10 MHz. Furthermore, $b_p=\lambda/\theta_p$ is longer at lower frequency, and therefore more baselines satisfy $b<2b_p$ and are included in the available data set. For a given direction, the total integration time for $b<2b_p$ at 3 MHz is about 3.3 times of the integration time at 10 MHz. Because of these two effect the overall SNR is actually better at lower frequency. 

However, note that in this analysis, we have not considered possible additional component of system noise which might arise at the low frequency end. Furthermore, here we have assumed that the same pixelization is used for these small regions as the global map. If the imaging is limited to a small region, more refined pixelization may be adopted, and correspondingly longer baselines can be effectively used in reconstruction, which would increase both the resolution and SNR. In this paper, we focus on full-sky imaging, and we would postpone detailed discussions on the imaging of a small region with higher resolution to a subsequent work.

\section{Error Analysis}
\label{sec:error}
As we noted in the previous section, the reconstructed map deviates from the true map due to both noise and reconstruction error, here we make a quantitative error analysis. First we introduce the covariance formalism, then we use the correlation coefficient between the input and reconstruction map, and the signal-to-noise ratio to quantify the quality of reconstruction. Based on these estimates, we discuss the choice of the regularization parameter and the sensitivity of the map.

\subsection{The Covariance Formalism}

\noindent 
\textbf{The thermal noise.}
Substituting Eq.~(\ref{eq:eigen_decomp_reg})-(\ref{eq:beam_eff}) into Eq.~(\ref{eq:mapmk}) with pure thermal noise $\matV=\bm{\eta}$ and null prior $\mats_p=0$, we obtain the covariance of error between pixels induced by thermal noise,
\begin{eqnarray}\label{eq:cov_pix}
\matSigma_n &=& \matB_{\rm eff}(\BTB + \matR)^{-1}\nonumber\\
&=& \matQ(\matW + \epsilon W_{\rm max}\matI)^{-1}\matW(\matW + \epsilon W_{\rm max}\matI)^{-1}\matQ^T\nonumber\\
&=& \frac{1}{\epsilon W_{\rm max}} \matB_{\rm eff}(\matI - \matB_{\rm eff})^T,
\end{eqnarray}
which is proportional to the effective beam weighted imperfection level. There are correlations between the pixel noise,  i.e. the non-diagonal entries in $\matSigma_n$ cannot be ignored, but the trace of $\matSigma_n$ can still reflect the overall thermal noise level in the reconstructed map. Since the trace of a matrix is the sum of its eigenvalues, from the second line of Eq.~(\ref{eq:cov_pix}), we find
\begin{equation}\label{eq:tr_ns}
{\rm Tr}(\matSigma_n) = \sum_i \frac{\matW_{ii}}{(\matW_{ii} + \epsilon W_{\rm max})^2}.
\end{equation}
For a fixed data set, the error in the map ${\rm Tr}(\matSigma_n)$ induced by the thermal noise is more suppressed with larger regularization parameter  $\epsilon$. \\

\noindent \textbf{The imperfect beam effect.}
We define $\matSigma_s\equiv(\matB_{\rm eff}-\matI)(\matB_{\rm eff}-\matI)^T$, and each diagonal entry of $\matSigma_s$ is
\begin{equation}\label{eq:sigma_s}
    (\matSigma_s)_{ii} = \sum_{k\neq i} (\matB_{\rm eff})_{ik}^2 + \left(1- (\matB_{\rm eff})_{ii} \right)^2,
\end{equation}
where the first term reflects the effect of sidelobes, and the second term reflect the effect of the normalization of the beam center.
Therefore, the trace of $\matSigma_s$ can be used to estimate the level of imperfection of $\matB_{\rm eff}$. Substituting Eq.~(\ref{eq:beam_eff}) into the definition of $\matSigma_s$ and following a procedure similar to that to derive Eq.~(\ref{eq:tr_ns}), we find
\begin{equation}\label{eq:tr_sl}
{\rm Tr}(\matSigma_s) = \sum_i \left(\frac{\epsilon W_{\rm max}}{\matW_{ii} + \epsilon W_{\rm max}}\right)^2.
\end{equation}
As the regularization parameter $\epsilon$ increases, the imperfect beam error ${\rm Tr}(\matSigma_s)$ increases. In the extreme case $\epsilon\gg 1$, $\matsnew\propto \BTV$, which is the dirty map. \\

\begin{figure}
\fig{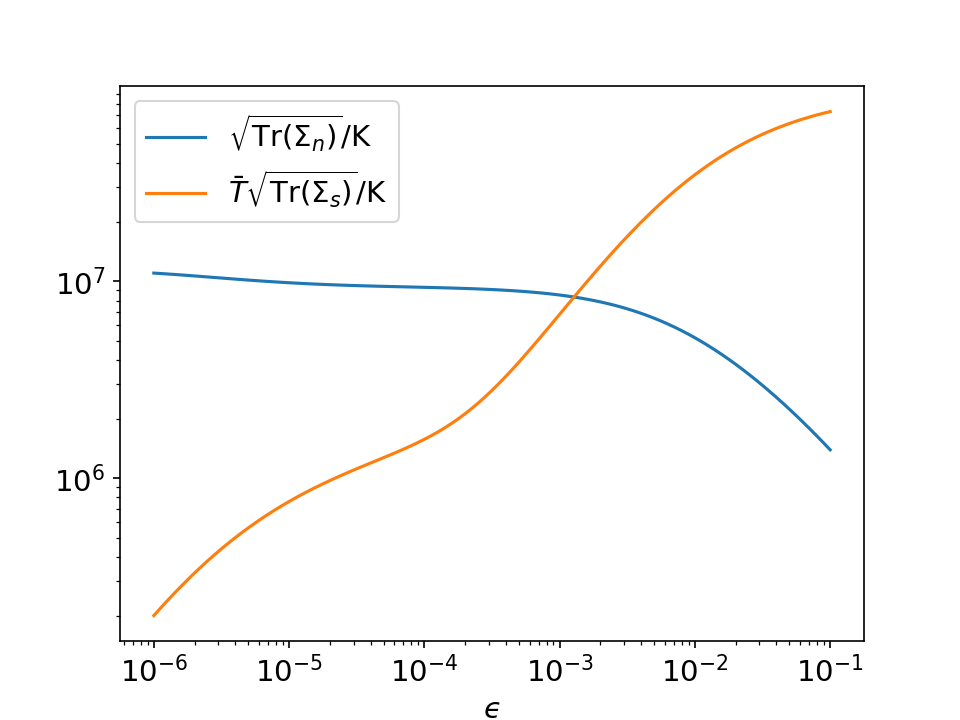}{0.48\textwidth}{}
\caption{The variation of the estimators of the error from the imperfect beam $\bar{T} {\rm Tr}(\matSigma_s)$ and that from the thermal noise ${\rm Tr}(\matSigma_n)$ for different $\epsilon$ at 10 MHz with baseline combination $b<2b_p$. 
}
\label{fig:sigma_trace_eps}
\end{figure}

In Figure \ref{fig:sigma_trace_eps}, we plot $\bar{T} {\rm Tr}(\matSigma_s)$ where $\bar{T}$ is the mean sky brightness temperature, and ${\rm Tr}(\matSigma_n)$, as a function of $\epsilon$ at 10 MHz with the baseline combination $b<2b_p$.
Note that ${\rm Tr}(\matSigma_s)$ does not have the same dimension as ${\rm Tr}(\matSigma_n)$.  
Now $\bar{T} {\rm Tr}(\matSigma_s)$ increases monotonically with increasing $\epsilon$, while ${\rm Tr}(\matSigma_n)$ is flat for small $\epsilon$, and decrease significantly for $\epsilon\gtrsim 10^{-3}$.
An `optimal' $\epsilon$ value can be found at the intersection of the two curves in Figure \ref{fig:sigma_trace_eps}, where the sum of the error is minimized. In the case plotted here, this is at $\epsilon \sim 2 \times 10^{-3}$. However,  we note that this result depends on many particular settings in the problem, e.g. the level of noise and signal, the amount of available data, etc., thus this particular number is only applicable for this particular case, it is not a universal number. In practice, it is probably necessary to try different values of $\epsilon$ and compare the results visually. Hopefully, good reconstruction maps consistent with each other could be obtained for a range of $\epsilon$ values.

In principle, the error induced by sub-pixel noise may also be computed using a similar formalism, which we outline in the Appendix. However, it turns out that this is computationally prohibitive. We therefore do not actually use it to compute the error. Instead, we use the correlation coefficient between the reconstructed image and input map, and the signal to noise ratio to evaluate the goodness of the reconstruction.

\begin{figure*}
\gridline{
\fig{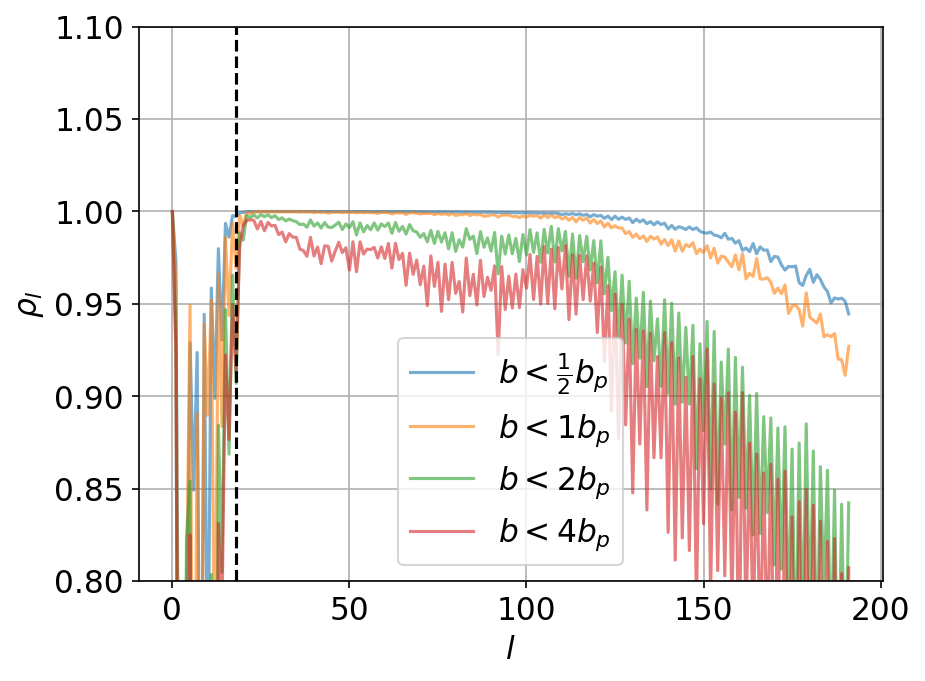}{0.45\textwidth}{(a) Pixel-center}
\fig{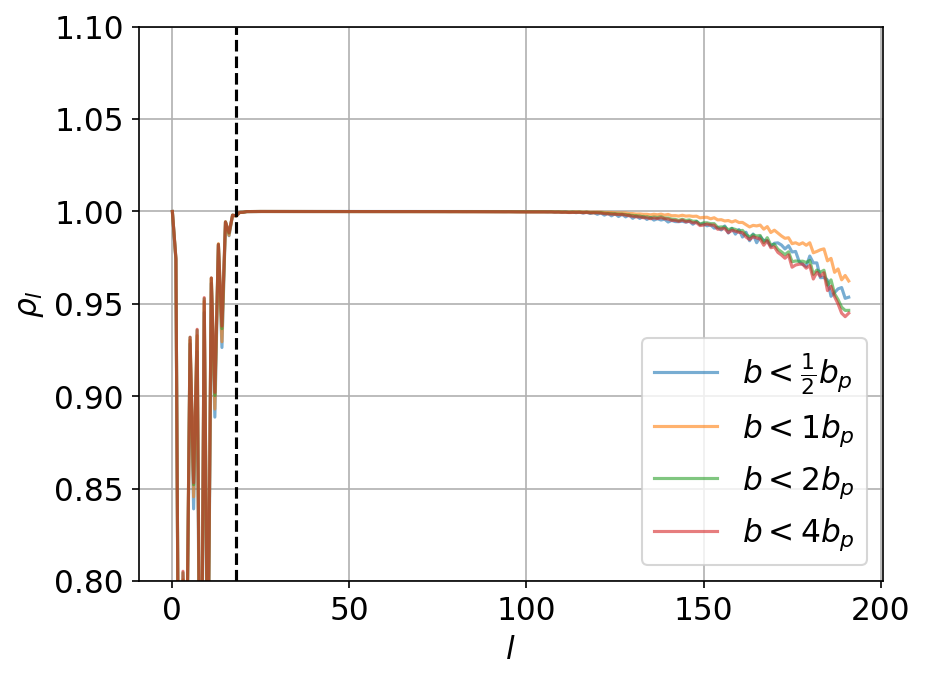}{0.45\textwidth}{(b) Pixel-averaging}
}
\caption{Comparison of the correlation coefficient $\rho_l$ at 10 MHz for the pixel-center (left) and pixel-averaging (right) methods, for different baseline combinations. The dashed vertical line marks $l=18$, below which $\rho_l$ drops.
}\label{fig:av_vs_ct_rho}
\end{figure*}

\subsection{Correlation Coefficient and Signal-to-Noise Ratio}

We can use the scale-dependent correlation coefficient $\rho_l$ to quantify the reconstruction error, which is defined as
\begin{equation}
    \rho_l = \frac{C_l^X}{\sqrt{C_l^{\rm in} C_l^{\rm re}}},
\end{equation}
where $C_l^{\rm in}$,$C_l^{\rm re}$ and $C_l^X$ are the angular power spectrum of the input high-resolution map, reconstructed map, and the cross-correlation respectively. In defining $\rho_l$ we do not include thermal noise. 

Besides the reconstruction error, in real measurement we will also have noise. We can use the Signal-to-Noise Ratio $\snrl$ in the spherical harmonic space to quantify this,
\begin{equation}
    \snrl = \frac{C_l^X}{\sqrt{C_l^{\rm in}C_l^{\rm ns}}},
\end{equation}
where $C_l^{\rm ns}$ is the angular power spectrum of the noise-only sky map. We use this expression to calculate the SNR instead of $C_l^{\rm re}/C_l^{\rm ns}$ because it does not contain the uncorrelated power caused by the reconstruction error and therefore can better reflect the SNR. 

We now use $\rho_l$ and $\snrl$ to investigate the errors in the reconstructed maps. First we look at $\rho_l$. 
In Figure \ref{fig:av_vs_ct_rho}, we show $\rho_l$ at 10 MHz for both the pixel-center and the pixel-averaging methods, and with different baseline combinations. Here the regularization parameter is set as $\epsilon=10^{-4}$, but we have checked that these results are not very sensitive to $\epsilon$. In Figure \ref{fig:av_vs_ct_rho} each of the combinations of baselines $b<b_p/2$, $b<b_p$, $b<2b_p$ and $b<4b_p$ at 10 MHz are shown.

When only baselines with $b<b_p/2$ are included, the full-sky satisfies the Nyquist sampling condition, and both the pixel-center and pixel average methods produce similar results.  For $18<l<130$, $\rho_l \sim 1$ which shows that on these scales the reconstruction is very well performed, while outside this range $\rho_l <1 $. 

For the cases where the longer baselines are included, i.e.,  $b<b_p$, $b<2b_p$ and $b<4b_p$, for the pixel-center method, $\rho_l$ drops well below 1 for the whole ranging, showing that the reconstruction is seriously affected by the sub-pixel noise. However, if the pixel-averaging method is employed, 
$\rho_l$ is not much changed and very similar to the case with $b<b_p/2$, only drops a little bit at $l \gtrsim 130$, showing that in this method the effect of the sub-pixel noise is effectively suppressed, allowing the longer baselines to be used without causing aliasing problem.  So below we focus on the pixel-averaging case.

Even for the pixel-averaging case, at $l\lesssim 18$, $\rho_l$ is significantly lower. This indicates relatively large reconstruction error on the large scales. 
This is consistent with our visual impression discussed in Section \ref{sec:global}, where we noted that the polar region is not well-reconstructed due to the lack of short baselines. We can also make a quantitative estimation: for our default array configuration, the shortest projected baseline at the polar region is $r_{\rm proj,min}=100\m\times\cos 30^\circ\approx 2.9\lambda$ at 10 MHz, and the corresponding angular scale is $l\sim 2\pi r_{\rm proj,min}/\lambda\approx 18$, which corresponds exactly to the scale where $\rho_l$ drops. 
Because of the lack of projected short baseline at the polar region, the response of the baselines to the large-scale modes with $l\lesssim 2\pi r_{\rm proj,min}/\lambda$ is incomplete, and the reconstruction of these modes is more vulnerable to noise.

\begin{figure}
\gridline{
\fig{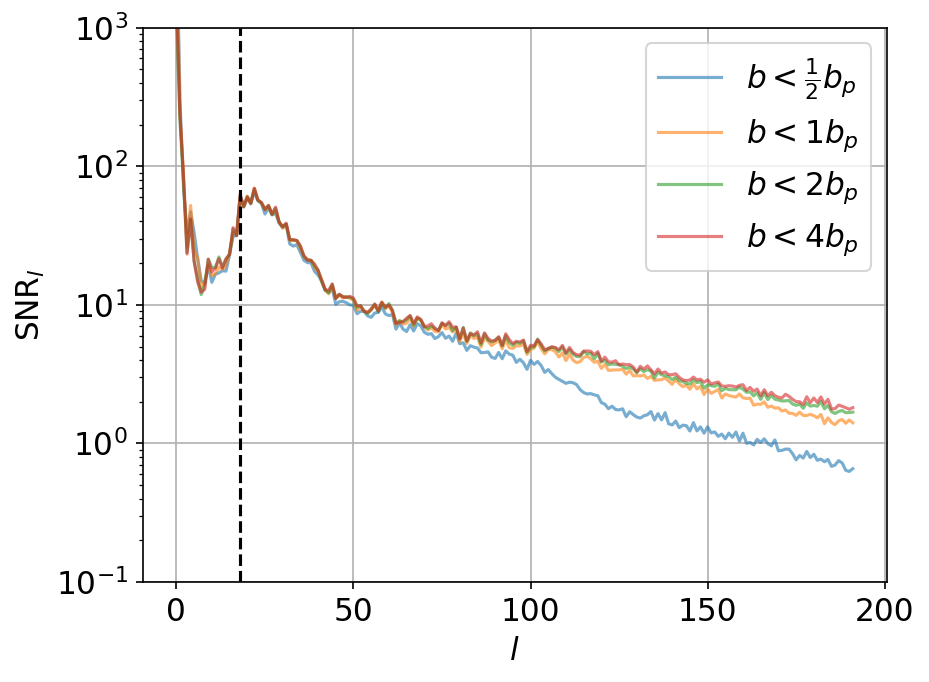}{0.45\textwidth}{}
}
\caption{The $\snrl$ for the pixel-averaging method at 10 MHz with different baseline combinations. Here we set $\epsilon=10^{-4}$ and the scale corresponding to $l=18$ is indicated by the dashed vertical line.
}\label{fig:snr_av_10MHz}
\end{figure}

\begin{figure}
\centering
\includegraphics[width=0.45\textwidth]{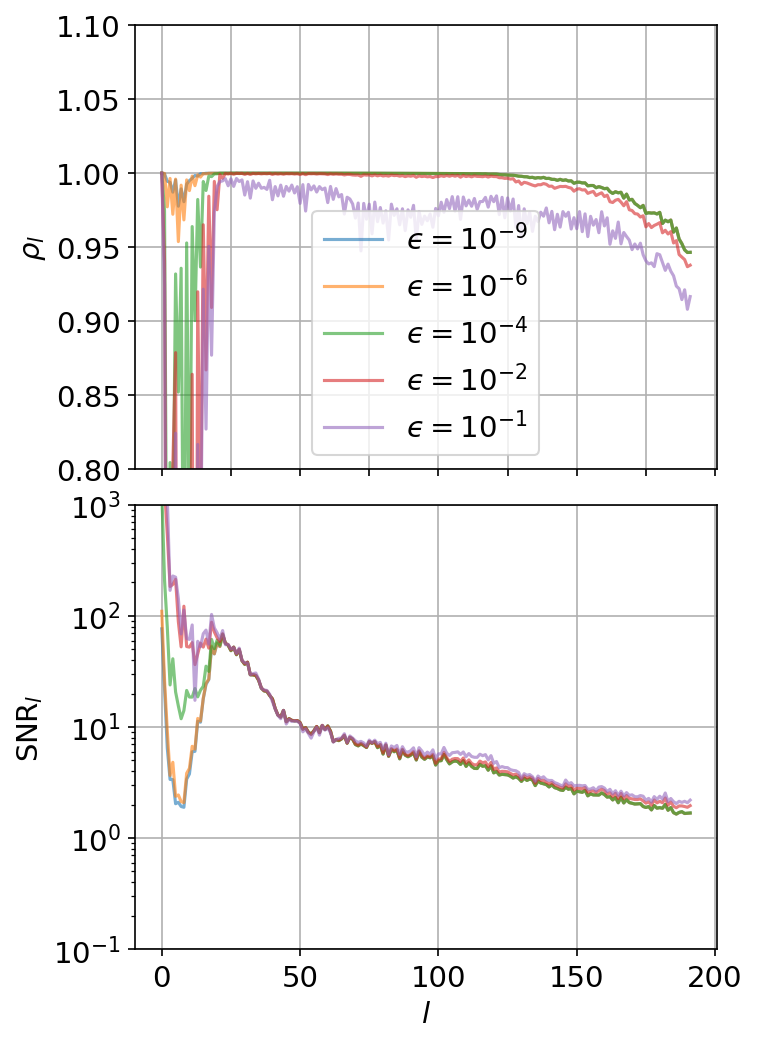}
\caption{The correlation coefficient $\rho_l$ and $\snrl$ as a function of $l$ for different regularization parameters $\epsilon$ at 10 MHz. }
\label{fig:eps_10MHz}
\end{figure}

In Figure \ref{fig:snr_av_10MHz} we show the $\snrl$ for the pixel-averaging method.  If only baselines with $b<b_p/2$ are included, the $\snrl$ drops more steeply at the high-$l$ end because the resolution of this baseline combination is limited.  When including longer baselines, the $\snrl$ improves significantly, especially at the high-$l$ end.

The results presented above show that the pixel-averaging method is robust against sub-pixel variation of the beam and produce high $\rho_l$ and good $\snrl$ when the longer baselines are included, while the pixel-center method fails when baselines longer than the Nyquist sampling are included. This justified our use of the pixel-averaging method as the standard method in our previous demonstrations. We also note that the $\rho_l$ and $\snrl$ curves do not differ much for the baseline combinations $b<2b_p$ and $b<4b_p$. This is because for baselines with $b>2b_p$, most entries of $\matB$ is damped to $\sim 0$ in the averaging process and make little contribution to the final map, or in other words, the scales larger than the pixel size $\theta_p$ are not sensitive to the baselines with $b>2b_p$, so in most results presented in this paper, the map-making is done only with the baselines $b<2b_p$ to save computation time.

\subsection{Choice of Regularization Parameter}
The results we obtained above are robust for a range of regularization parameter $\epsilon$, but the choice of the value of  $\epsilon$ does affect the reconstruction result. In Figure \ref{fig:eps_10MHz}, we plot  $\rho_l$ and $\snrl$ with different $\epsilon$ values for the baseline combination $b<2b_p$.
At the high-$l$ end, as long as $\epsilon<10^{-2}$, both $\rho_l$ and $\snrl$ are insensitive to the change of the regularization parameter. Only when $\epsilon\gtrsim 10^{-2}$, $\rho_l$ drops below 1 at all scales, indicating large reconstruction error.

At the low-$l$ end, both $\rho_l$ and $\snrl$ vary significantly with $\epsilon$. For large $\epsilon$, the thermal noise-induced error is expected to be suppressed according to Eq.~(\ref{eq:tr_ns}), so $\snrl$ would be larger, but at the same time the imperfect beam effect would be larger, thus decreasing $\rho_l$ according to Eq.~(\ref{eq:tr_sl}). The result confirms these expectations. However, even for the $\epsilon=10^{-9}$ case, where the regularization can be ignored, $\rho_l$ does not reach 1 at low-$l$ end. Both the regularization error and sub-pixel noise contribute to the distortion which causes $\rho_l$ to drop at low-$l$ end. 

Thus, when choosing  $\epsilon$, there is a trade-off in optimizing for $\rho_l$ and $\snrl$. We have chosen an intermediate $\epsilon=10^{-4}$ as our fiducial case. In real observations, other sources of noise may also be present, such as 1/f noise or RFI. Using a higher $\epsilon$ will make the results more robust against such errors. 

\subsection{Sensitivity}

\begin{figure}
\gridline{
\fig{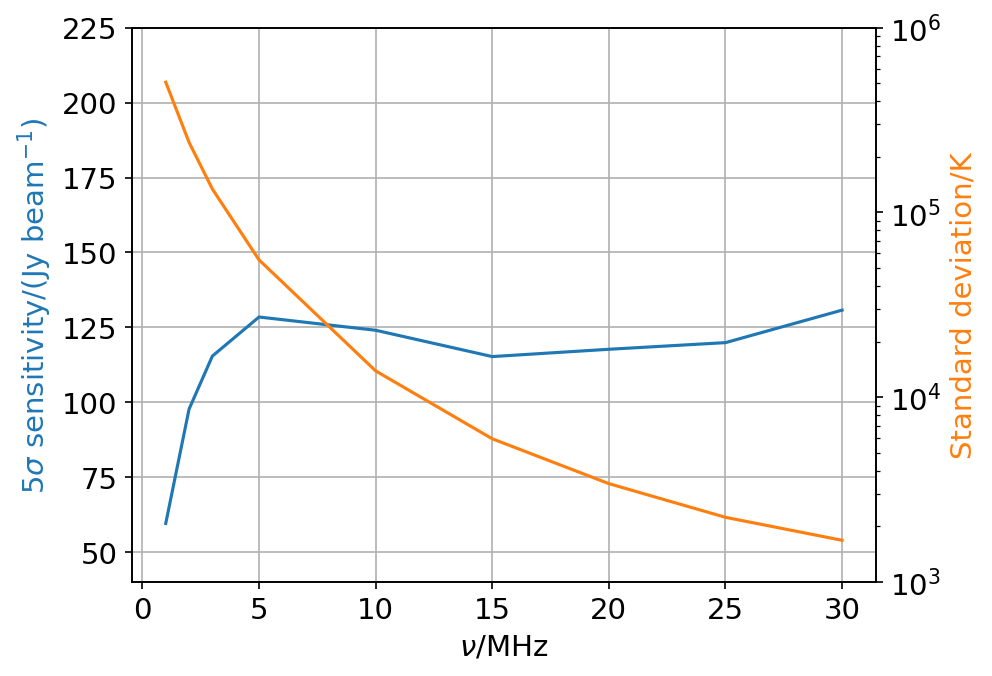}{0.45\textwidth}{}
}
\caption{The $5\sigma$ point source sensitivity (left axis, blue line) estimated from the noise-only map and the standard deviation of the same map (right axis, orange line) for baseline combination $b<2b_p$ and $\epsilon=10^{-3}$ after post-processing and removal of large-scale component. }
\label{fig:ps_sens}
\end{figure}

We can estimate the point-source sensitivity in this map. By calculating the covariance given by Eq.~(\ref{eq:cov_pix}), we find the standard deviation of the noise is quite uniform across the whole sky, with typical fluctuation of $\sim 10\%$, while different projected $uv$ coverage for different sky area influence the correlation between pixels. After post-processing and removal of large scale component, the $5\sigma$ point-source sensitivity is estimated from the standard deviation of the noise-only map. We convert the result to the unit of $\Jy/{\rm beam}$ using the estimated beam size, and plot this by the blue line in Figure \ref{fig:ps_sens} for the baseline combination $b<2b_p$. We also show the standard deviation ($1\sigma$) of the noise-only map as the orange line in the same figure. Here we choose a slightly higher $\epsilon=10^{-3}$ to suppress the correlated noise in polar regions, which improves the sensitivity for observations at $\nu>10 \MHz$ while preserving a decent correlation coefficient. We also compared this result with the sensitivity estimated solely from equatorial regions which is quite insensitive to regularization, and we found that they are consistent. The sensitivity is quite flat for $\nu\geq 5\MHz$ due to the competition between higher sky brightness temperature and more integration time toward low-frequency end, while for $\nu<5\MHz$, the sensitivity increases quickly toward low-frequency end due to suppression of sky brightness temperature caused by absorption. Note however that here we assume the noise is essentially dominated by and scales with the sky temperature at low frequencies. If there are additional noise component, or the noise does not scale with the sky temperature at low frequencies, the situation may be different. In those cases, the result would depend on the specific noise model.  

\section{Conclusion}
\label{sec:conclusion}
In this work, we studied the global map-making of the DSL project at a HealPix pixelization with $\nside=64$, corresponding to an the angular scale of $\sim\ 1.3^\circ$ after post-processing. As the DSL have baselines which probe finer angular scales, the anisotropy on smaller scales which we call the sub-pixel noise may affect the map-making. We find that if we construct the beam matrix $\matB$ by simply sampling at the center of each pixel, it would suffer from aliasing of the sub-pixel noise when baselines beyond the Nyquist sampling are included, but this aliasing effect can be mitigated by employing a simple pixel-averaging method. We also find that including all baselines with $b<2b_p$ is sufficient to make map for given pixel size $\theta_p$, while including longer baselines do not improve the quality of the map. 

The map-making quality at 10 MHz suffers from the low SNR at small-scales, while on large scales some modes are lost due to the lack of short projected baselines at the polar region. At lower frequency, e.g. 3 MHz, the sky are better reconstructed on large scales, due to the inclusion of more baselines and better $uvw$ coverage. We also investigate the impact of the regularization parameter $\epsilon$, and find that its value can be chosen as a trade-off between the reconstruction error and SNR.

In the present work we studied the imaging in an idealized model. We considered some practical issues and systematics in this study, but we also neglected or ignored a number of systematic effects, such as the baseline determination and instrument calibration error, the impact of variable sources such as the Sun and planets, the reflection of the Moon. These effects will be further investigated in subsequent papers. Furthermore, in the actual operation there may also be other practical issues which limit the data collection. Nevertheless, the results of this study can illustrate some basic characters of the synthesis imaging of the DSL mission. 

\begin{acknowledgements}
We thank Prof. Li Deng (NSSC) and Dr. Yuan Shi (SJTU) for discussions. This work is supported by National Key R\&D Program of China No. 2022YFF0504300,  China's  Space Origins Exploration Program No. GJ11010401, and the National Natural Science Foundation of China (NSFC) grants No. 12361141814, 11973047, 12273070.

\end{acknowledgements}

\appendix

\section{Error induced by sub-pixel noise.}
To investigate errors arising from the finite resolution of the pixelization, we work in the spherical harmonic space. The visibility can be transformed as \citep{2014ApJ...781...57S,Zhang:2016whm},
\begin{equation}
    V_\alpha = \sum_{l, m} s_{lm} (b_\alpha^*)_{lm}
\end{equation}
where $s_{lm}$ and $(b_\alpha)_{lm}$ are the spherical harmonic coefficients for the sky and for the beam respectively. We can decompose $(b_\alpha)_{lm}$ into two parts: $(b_\alpha^0)_{lm}$ which is the spherical harmonic transform of the model $\matB$, and $(\Delta b_\alpha)_{lm}=(b_\alpha)_{lm}-(b_\alpha^0)_{lm}$, which is the deviation of the model from the true beam due to the finite resolution. $(\Delta b_\alpha)_{lm}$ introduced a `noise-like' term in the visibility,
\begin{equation}
    n_\alpha^b = \sum_{l, m}s_{lm}(\Delta b_\alpha)_{lm}.
\end{equation}
If $\matB$ correctly captures the large-scale components of $(b_\alpha)_{lm}$, $(\Delta b_\alpha)_{lm}$ would be significant only for high $l$.
We assume that 
$\langle s_{lm} s^*_{l'm'}\rangle=C_l\delta_{mm'}\delta_{ll'}$ for high $l$, then the covariance of $n_\alpha^b$ can be simplified to
\begin{eqnarray}\label{eq:cov_subpix_est}
    (\matN_b)_{\alpha\alpha'}
    &=& \langle n_\alpha^b n_{\alpha'}^{b*} \rangle\nonumber\\
    &=& \sum_{l, m} (\Delta b_\alpha^*)_{lm} (\Delta b_{\alpha'})_{lm} C_l,
\end{eqnarray}
and once we assume some angular power spectrum $C_l$, the covariance can be evaluated. The larger  small-scale power $C_l$ has the larger sub-pixel noise is induced, and the larger error we would have in the reconstructed maps.

\begin{figure}
\fig{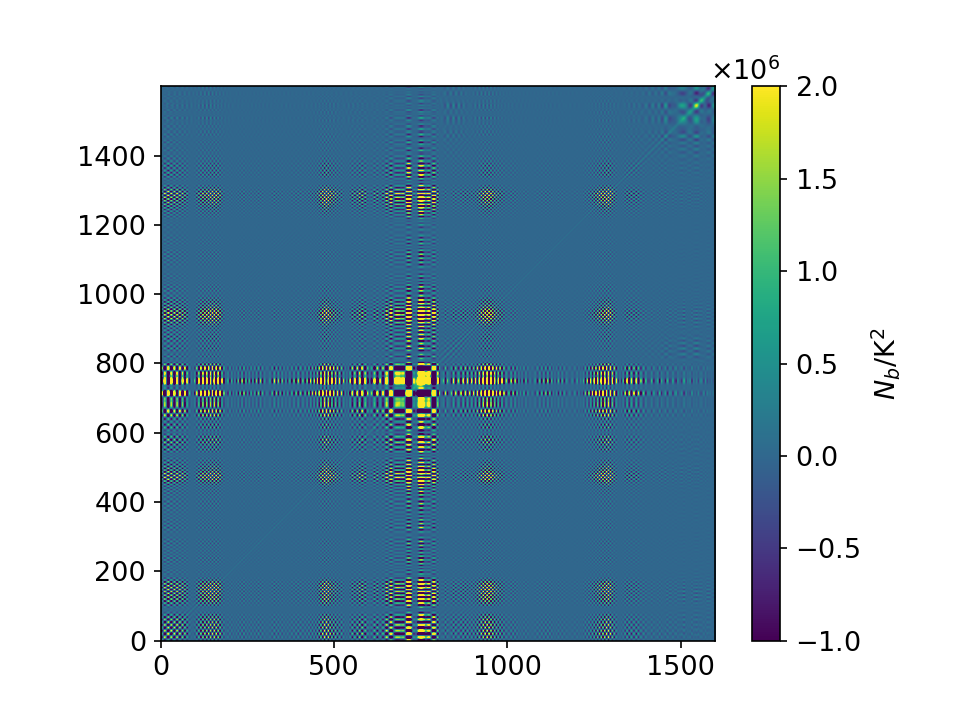}{0.48\textwidth}{(a) Pixel-center}
\fig{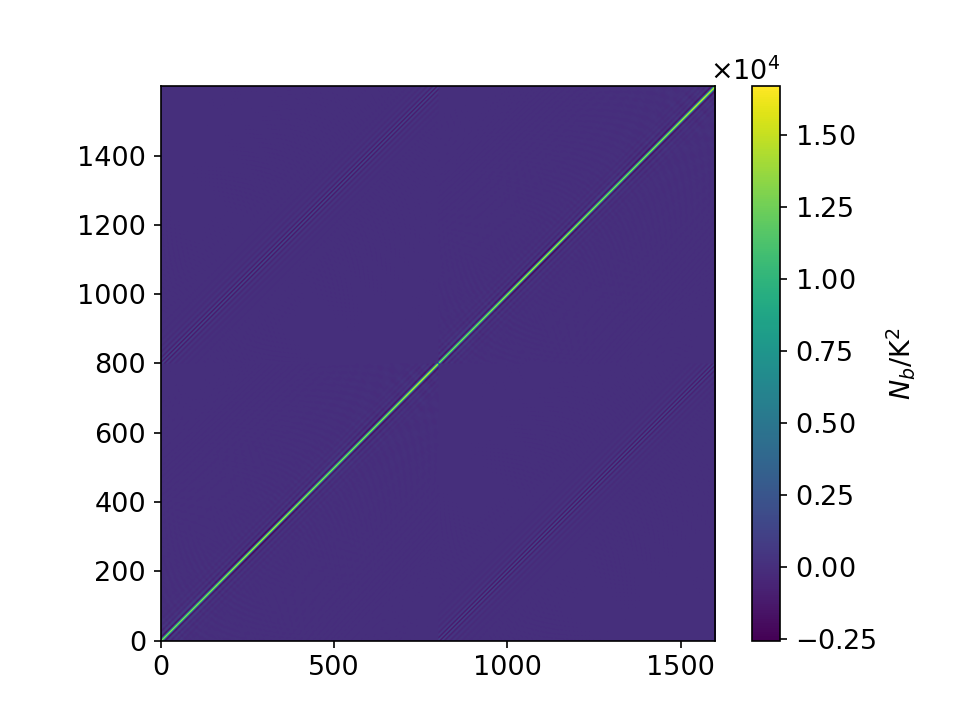}{0.48\textwidth}{(b) Pixel-averaging}
\caption{Noise covariance induced by the sub-pixel noise $\matN_b$ estimated using Eq.~ (\ref{eq:cov_subpix_est}) for the same baseline and time points as in Figure \ref{fig:vis_av_vs_ct} at 10 MHz for both pixel-center (left panel) and pixel-averaging right panel) methods. 
}\label{fig:cov_Nb}
\end{figure}

In Figure \ref{fig:cov_Nb}, we show $\matN_b$ estimated using Eq.~(\ref{eq:cov_subpix_est}) for a baseline with $b=b_p$ at 10 MHz, same as the one shown in Figure \ref{fig:vis_av_vs_ct}, for both the pixel-center method and the pixel-averaging method. The figure shows 800 time points, and the real and imaginary parts are concatenated, giving a size of $(1600\times 1600)$. The sub-pixel noise is much higher and more strongly correlated for the pixel-center method, while the pixel-averaging method significantly alleviates both the amplitude and the correlation of the sub-pixel noise. We have also compared the estimated visibility covariance $\matN_b$ with the residual in visibility caused by sub-pixel noise and found reasonable consistency.

The covariance in the reconstructed sky map caused by $\matN_b$ is
\begin{eqnarray}\label{eq:cov_map_subpix_est}
    \matSigma_b =&& (\BTB+\matR)^{-1} \BT \matN_b \nonumber\\
    &&\matN^{-1}\matB(\BTB+\matR)^{-1}\nonumber\\
    =&& \matQ \mathbf{\Lambda}_b \matQ^T
\end{eqnarray}
where the $\mathbf{\Lambda}_b$ in the last line is
\begin{equation}\label{eq:cov_Nb}
    (\mathbf{\Lambda}_b)_{ij} = \frac{(\matN_b)_{ij}\sqrt{\matW_{ii}\matW_{jj}}}{ (\matW_{ii}+\epsilon W_{\rm max}) (\matW_{jj}+\epsilon W_{\rm max}) }.
\end{equation}
In deriving this, we have used the fact that for diagonal noise matrix $\matN$, $\matB^T \matN^{-\frac{1}{2}} = \matQ \matW^{\frac{1}{2}}$, 
where 
$(\matN^{-\frac{1}{2}})_{ij} = \delta_{ij}/\sqrt{\matN_{ii}}$ and $(\matW^{\frac{1}{2}})_{ij} = \delta_{ij} \sqrt{\matW_{ii}}$.

However, estimating the error in this way requires inversion of matrix, which takes a large amount of computation.

\bibliography{DSL_low_resol}{}

\begin{thebibliography}{}
\expandafter\ifx\csname natexlab\endcsname\relax\def\natexlab#1{#1}\fi
\providecommand{\url}[1]{\href{#1}{#1}}
\providecommand{\dodoi}[1]{doi:~\href{http://doi.org/#1}{\nolinkurl{#1}}}
\providecommand{\doeprint}[1]{\href{http://ascl.net/#1}{\nolinkurl{http://ascl.net/#1}}}
\providecommand{\doarXiv}[1]{\href{https://arxiv.org/abs/#1}{\nolinkurl{https://arxiv.org/abs/#1}}}

\bibitem[{{Artuc} \& {de Lera Acedo}(2025)}]{Artuc2024}
{Artuc}, K., \& {de Lera Acedo}, E. 2025, RAS Techniques and Instruments, 4,
  rzae061, \dodoi{10.1093/rasti/rzae061}

\bibitem[{{Bale} {et~al.}(2023){Bale}, {Bassett}, {Burns}, {Dorigo Jones},
  {Goetz}, {Hellum-Bye}, {Hermann}, {Hibbard}, {Maksimovic}, {McLean},
  {Monsalve}, {O'Connor}, {Parsons}, {Pulupa}, {Pund}, {Rapetti}, {Rotermund},
  {Saliwanchik}, {Slosar}, {Sundkvist}, \& {Suzuki}}]{2023arXiv230110345B}
{Bale}, S.~D., {Bassett}, N., {Burns}, J.~O., {et~al.} 2023, arXiv e-prints,
  arXiv:2301.10345, \dodoi{10.48550/arXiv.2301.10345}

\bibitem[{Bandyopadhyay {et~al.}(2021)Bandyopadhyay, Mcgarey, Goel, Rafizadeh,
  Delapierre, Arya, Lazio, Goldsmith, Chahat, Stoica, Quadrelli, Nesnas, Jenks,
  \& Hallinan}]{LCRT2021}
Bandyopadhyay, S., Mcgarey, P., Goel, A., {et~al.} 2021, in 2021 IEEE Aerospace
  Conference (50100), 1--25, \dodoi{10.1109/AERO50100.2021.9438165}

\bibitem[{{Burns}(2021)}]{2021RSPTA.37990564B}
{Burns}, J.~O. 2021, Philosophical Transactions of the Royal Society of London
  Series A, 379, 20190564, \dodoi{10.1098/rsta.2019.0564}

\bibitem[{Chen {et~al.}(2018)Chen, Aminaei, Gurvits, Wolt, Pourshaghaghi, Yan,
  \& Falcke}]{chen2018antenna}
Chen, L., Aminaei, A., Gurvits, L.~I., {et~al.} 2018, Experimental Astronomy,
  45, 231

\bibitem[{Chen {et~al.}(2020)Chen, Yan, Deng, Wu, Wu, Xu, \&
  Zhou}]{Chen:2020lok}
Chen, X., Yan, J., Deng, L., {et~al.} 2020, Phil. Trans. Roy. Soc. Lond. A,
  379, 20190566, \dodoi{10.1098/rsta.2019.0566}

\bibitem[{{Chen} {et~al.}(2019){Chen}, {Burns}, {Koopmans}, {Rothkaehi},
  {Silk}, {Wu}, {Boonstra}, {Cecconi}, {Chiang}, {Chen}, {Deng}, {Falanga},
  {Falcke}, {Fan}, {Fang}, {Fialkov}, {Gurvits}, {Ji}, {Kasper}, {Li}, {Mao},
  {Mckinley}, {Monsalve}, {Peterson}, {Ping}, {Subrahmanyan}, {Vedantham},
  {Klein Wolt}, {Wu}, {Xu}, {Yan}, \& {Yue}}]{Chen2019}
{Chen}, X., {Burns}, J., {Koopmans}, L., {et~al.} 2019, arXiv e-prints,
  arXiv:1907.10853.
\newblock \doarXiv{1907.10853}

\bibitem[{Chen {et~al.}(2023)Chen, Yan, Xu, Deng, Wu, Wu, Zhou, Zhang, Zhu,
  Yang, \& Wu}]{chen2023}
Chen, X., Yan, J., Xu, Y., {et~al.} 2023, Chinese Journal of Space Science, 43,
  43

\bibitem[{Chen {et~al.}(2024)Chen, Gao, Wu, Zhang, Wang, Liu, Zou, Deng, Gong,
  He, {et~al.}}]{chen2024large}
Chen, X., Gao, F., Wu, F., {et~al.} 2024, Philosophical Transactions of the
  Royal Society A, 382, 20230094

\bibitem[{Cong {et~al.}(2024)Cong, Yue, Xu, Deng, Zhang, \&
  Chen}]{cong2024loop}
Cong, Y., Yue, B., Xu, Y., {et~al.} 2024, arXiv preprint arXiv:2501.00431

\bibitem[{{Cong} {et~al.}(2021){Cong}, {Yue}, {Xu}, {Huang}, {Zuo}, \&
  {Chen}}]{2021ApJ...914..128C}
{Cong}, Y., {Yue}, B., {Xu}, Y., {et~al.} 2021, \apj, 914, 128,
  \dodoi{10.3847/1538-4357/abf55c}

\bibitem[{{Cong} {et~al.}(2022){Cong}, {Yue}, {Xu}, {Shi}, \&
  {Chen}}]{2022ApJ...940..180C}
{Cong}, Y., {Yue}, B., {Xu}, Y., {Shi}, Y., \& {Chen}, X. 2022, \apj, 940, 180,
  \dodoi{10.3847/1538-4357/ac9df7}

\bibitem[{{Cornwell} {et~al.}(2008){Cornwell}, {Golap}, \&
  {Bhatnagar}}]{2008ISTSP...2..647C}
{Cornwell}, T.~J., {Golap}, K., \& {Bhatnagar}, S. 2008, IEEE Journal of
  Selected Topics in Signal Processing, 2, 647,
  \dodoi{10.1109/JSTSP.2008.2005290}

\bibitem[{{Cornwell} \& {Perley}(1992)}]{1992A&A...261..353C}
{Cornwell}, T.~J., \& {Perley}, R.~A. 1992, \aap, 261, 353

\bibitem[{{Cornwell} {et~al.}(2012){Cornwell}, {Voronkov}, \&
  {Humphreys}}]{2012SPIE.8500E..0LC}
{Cornwell}, T.~J., {Voronkov}, M.~A., \& {Humphreys}, B. 2012, in Society of
  Photo-Optical Instrumentation Engineers (SPIE) Conference Series, Vol. 8500,
  Image Reconstruction from Incomplete Data VII, ed. P.~J. {Bones}, M.~A.
  {Fiddy}, \& R.~P. {Millane}, 85000L, \dodoi{10.1117/12.929336}

\bibitem[{de~Oliveira-Costa {et~al.}(2008)de~Oliveira-Costa, Tegmark, Gaensler,
  Jonas, Landecker, \& Reich}]{deOliveiraCosta:2008pb}
de~Oliveira-Costa, A., Tegmark, M., Gaensler, B.~M., {et~al.} 2008, Mon. Not.
  Roy. Astron. Soc., 388, 247, \dodoi{10.1111/j.1365-2966.2008.13376.x}

\bibitem[{{Eastwood} {et~al.}(2018){Eastwood}, {Anderson}, {Monroe},
  {Hallinan}, {Barsdell}, {Bourke}, {Clark}, {Ellingson}, {Dowell}, {Garsden},
  {Greenhill}, {Hartman}, {Kocz}, {Lazio}, {Price}, {Schinzel}, {Taylor},
  {Vedantham}, {Wang}, \& {Woody}}]{eastwood2018}
{Eastwood}, M.~W., {Anderson}, M.~M., {Monroe}, R.~M., {et~al.} 2018, \aj, 156,
  32, \dodoi{10.3847/1538-3881/aac721}

\bibitem[{{Evans}(1969)}]{Evans_1969}
{Evans}, J.~V. 1969, \araa, 7, 201, \dodoi{10.1146/annurev.aa.07.090169.001221}

\bibitem[{{Franzen} {et~al.}(2019){Franzen}, {Vernstrom}, {Jackson},
  {Hurley-Walker}, {Ekers}, {Heald}, {Seymour}, \&
  {White}}]{2019PASA...36....4F}
{Franzen}, T.~M.~O., {Vernstrom}, T., {Jackson}, C.~A., {et~al.} 2019, \pasa,
  36, e004, \dodoi{10.1017/pasa.2018.52}

\bibitem[{Golub {et~al.}(1979)Golub, Heath, \& Wahba}]{golub1979}
Golub, G.~H., Heath, M., \& Wahba, G. 1979, Technometrics, 21, 215.
\newblock \url{http://www.jstor.org/stable/1268518}

\bibitem[{{G{\'o}rski} {et~al.}(2005){G{\'o}rski}, {Hivon}, {Banday},
  {Wandelt}, {Hansen}, {Reinecke}, \& {Bartelmann}}]{2005ApJ...622..759G}
{G{\'o}rski}, K.~M., {Hivon}, E., {Banday}, A.~J., {et~al.} 2005, \apj, 622,
  759, \dodoi{10.1086/427976}

\bibitem[{{Hibbard} {et~al.}(2025){Hibbard}, {Burns}, {MacDowall},
  {Gopalswamy}, {Boardsen}, {Farrell}, {Bradley}, {Schulszas}, {Dorigo Jones},
  {Rapetti}, \& {Turner}}]{Hibbard2025}
{Hibbard}, J.~J., {Burns}, J.~O., {MacDowall}, R., {et~al.} 2025, arXiv
  e-prints, arXiv:2503.09842, \dodoi{10.48550/arXiv.2503.09842}

\bibitem[{{Huang} {et~al.}(2018){Huang}, {Sun}, {Zuo}, {Wu}, {Xu}, {Yue},
  {Ansari}, \& {Chen}}]{Huang2018}
{Huang}, Q., {Sun}, S., {Zuo}, S., {et~al.} 2018, \aj, 156, 43,
  \dodoi{10.3847/1538-3881/aac6c6}

\bibitem[{{Huang} {et~al.}(2019){Huang}, {Wu}, \& {Chen}}]{Huang2019}
{Huang}, Q., {Wu}, F., \& {Chen}, X. 2019, Science China Physics, Mechanics,
  and Astronomy, 62, 989511, \dodoi{10.1007/s11433-018-9333-1}

\bibitem[{{Klein Wolt} {et~al.}(2024){Klein Wolt}, {Falcke}, \&
  {Koopmans}}]{2024AAS...24326401K}
{Klein Wolt}, M., {Falcke}, H., \& {Koopmans}, L. 2024, in American
  Astronomical Society Meeting Abstracts, Vol. 243, American Astronomical
  Society Meeting Abstracts \#243, 264.01

\bibitem[{{Kriele} {et~al.}(2022){Kriele}, {Wayth}, {Bentum}, {Juswardy}, \&
  {Trott}}]{eda2022}
{Kriele}, M.~A., {Wayth}, R.~B., {Bentum}, M.~J., {Juswardy}, B., \& {Trott},
  C.~M. 2022, \pasa, 39, e017, \dodoi{10.1017/pasa.2022.2}

\bibitem[{{Mondal} \& {Barkana}(2023)}]{2023NatAs...7.1025M}
{Mondal}, R., \& {Barkana}, R. 2023, Nature Astronomy, 7, 1025,
  \dodoi{10.1038/s41550-023-02057-y}

\bibitem[{{Plice} {et~al.}(2017){Plice}, {Galal}, \&
  {Burns}}]{2017arXiv170200286P}
{Plice}, L., {Galal}, K., \& {Burns}, J.~O. 2017, arXiv e-prints,
  arXiv:1702.00286, \dodoi{10.48550/arXiv.1702.00286}

\bibitem[{{Remazeilles} {et~al.}(2015){Remazeilles}, {Dickinson}, {Banday},
  {Bigot-Sazy}, \& {Ghosh}}]{2015MNRAS.451.4311R}
{Remazeilles}, M., {Dickinson}, C., {Banday}, A.~J., {Bigot-Sazy}, M.~A., \&
  {Ghosh}, T. 2015, \mnras, 451, 4311, \dodoi{10.1093/mnras/stv1274}

\bibitem[{{Sathyanarayana Rao} {et~al.}(2023){Sathyanarayana Rao}, {Singh},
  {K.~S.}, {B.~S.}, {Sathish}, {Somashekar}, {Agaram}, {Kavitha},
  {Vishwapriya}, {Anand}, {Udaya Shankar}, \& {Seetha}}]{2023ExA....56..741S}
{Sathyanarayana Rao}, M., {Singh}, S., {K.~S.}, S., {et~al.} 2023, Experimental
  Astronomy, 56, 741, \dodoi{10.1007/s10686-023-09909-5}

\bibitem[{{Shaw} {et~al.}(2014){Shaw}, {Sigurdson}, {Pen}, {Stebbins}, \&
  {Sitwell}}]{2014ApJ...781...57S}
{Shaw}, J.~R., {Sigurdson}, K., {Pen}, U.-L., {Stebbins}, A., \& {Sitwell}, M.
  2014, \apj, 781, 57, \dodoi{10.1088/0004-637X/781/2/57}

\bibitem[{Shi {et~al.}(2022)Shi, Deng, Xu, Wu, Yan, \& Chen}]{Shi:2022zdx}
Shi, Y., Deng, F., Xu, Y., {et~al.} 2022, Astrophys. J., 929, 32,
  \dodoi{10.3847/1538-4357/ac5965}

\bibitem[{{Shi} {et~al.}(2022){Shi}, {Xu}, {Deng}, {Wu}, {Wu}, {Huang}, {Zuo},
  {Yan}, \& {Chen}}]{Shi:2022xdw}
{Shi}, Y., {Xu}, Y., {Deng}, L., {et~al.} 2022, \mnras, 510, 3046,
  \dodoi{10.1093/mnras/stab3623}

\bibitem[{{Slob} {et~al.}(2022){Slob}, {Callingham}, {R{\"o}ttgering},
  {Williams}, {Duncan}, {de Gasperin}, {Hardcastle}, \&
  {Miley}}]{2022A&A...668A.186S}
{Slob}, M.~M., {Callingham}, J.~R., {R{\"o}ttgering}, H.~J.~A., {et~al.} 2022,
  \aap, 668, A186, \dodoi{10.1051/0004-6361/202244651}

\bibitem[{Tanti \& Datta(2021)}]{SEAMS2021}
Tanti, H.~A., \& Datta, A. 2021, in 2021 IEEE Indian Conference on Antennas and
  Propagation (InCAP), 312--315, \dodoi{10.1109/InCAP52216.2021.9726500}

\bibitem[{{Vecchio} {et~al.}(2021){Vecchio}, {Bentum}, {Falcke}, {Boonstra},
  {Ping}, {Chen}, {Klein-Wolt}, {Brinkerink}, {Rotteveel}, {Pourshaghaghi},
  {Karapakula}, {Ruiter}, \& {Bertels}}]{Vecchio2021}
{Vecchio}, A., {Bentum}, M., {Falcke}, H., {et~al.} 2021, in 43rd COSPAR
  Scientific Assembly. Held 28 January - 4 February, Vol.~43, 1525

\bibitem[{Wu {et~al.}(2024)Wu, Xu, Zhu, Zhang, He, Sun, Zhou, Zhou, Xu, Zhou,
  Zhang, Yan, Yan, Li, Deng, Yujie, Suo, Zhang, Zhang, \& Chen}]{wu2024}
Wu, F., Xu, j., Zhu, J., {et~al.} 2024, Chinese journal of radio science, 39,
  589

\bibitem[{{Yan} {et~al.}(2023){Yan}, {Wu}, {Gurvits}, {Wu}, {Deng}, {Zhao},
  {Zhou}, {Lan}, {Fan}, {Yi}, {Yang}, {Yang}, {Wei}, {Guo}, {Qiu}, {Wu}, {Hu},
  {Chen}, {Rothkaehl}, \& {Morawski}}]{2023ExA....56..333Y}
{Yan}, J., {Wu}, J., {Gurvits}, L.~I., {et~al.} 2023, Experimental Astronomy,
  56, 333, \dodoi{10.1007/s10686-022-09887-0}

\bibitem[{Yu {et~al.}(2024)Yu, Zuo, Wu, Wang, \& Chen}]{Yu:2023idm}
Yu, K., Zuo, S., Wu, F., Wang, Y., \& Chen, X. 2024, Res. Astron. Astrophys.,
  24, 025002, \dodoi{10.1088/1674-4527/ad1223}

\bibitem[{Zhang {et~al.}(2016)Zhang, Ansari, Chen, Campagne, Magneville, \&
  Wu}]{Zhang:2016whm}
Zhang, J., Ansari, R., Chen, X., {et~al.} 2016, Mon. Not. Roy. Astron. Soc.,
  461, 1950, \dodoi{10.1093/mnras/stw1458}

\bibitem[{{Zhang} {et~al.}(2012){Zhang}, {Gray}, {Su}, {Li}, {Landecker},
  {Zhang}, \& {Li}}]{Zhang_2012}
{Zhang}, X.-Z., {Gray}, A., {Su}, Y., {et~al.} 2012, Research in Astronomy and
  Astrophysics, 12, 1297, \dodoi{10.1088/1674-4527/12/9/010}

\bibitem[{{Zheng} {et~al.}(2017{\natexlab{a}}){Zheng}, {Tegmark}, {Dillon},
  {Liu}, {Neben}, {Tribiano}, {Bradley}, {Buza}, {Ewall-Wice}, {Gharibyan},
  {Hickish}, {Kunz}, {Losh}, {Lutomirski}, {Morgan}, {Narayanan}, {Perko},
  {Rosner}, {Sanchez}, {Schutz}, {Valdez}, {Villasenor}, {Yang}, {Zarb Adami},
  {Zelko}, \& {Zheng}}]{Zheng2017}
{Zheng}, H., {Tegmark}, M., {Dillon}, J.~S., {et~al.} 2017{\natexlab{a}},
  \mnras, 465, 2901, \dodoi{10.1093/mnras/stw2910}

\bibitem[{{Zheng} {et~al.}(2017{\natexlab{b}}){Zheng}, {Tegmark}, {Dillon},
  {Liu}, {Neben}, {Tribiano}, {Bradley}, {Buza}, {Ewall-Wice}, {Gharibyan},
  {Hickish}, {Kunz}, {Losh}, {Lutomirski}, {Morgan}, {Narayanan}, {Perko},
  {Rosner}, {Sanchez}, {Schutz}, {Valdez}, {Villasenor}, {Yang}, {Zarb Adami},
  {Zelko}, \& {Zheng}}]{2017MNRAS.465.2901Z}
---. 2017{\natexlab{b}}, \mnras, 465, 2901, \dodoi{10.1093/mnras/stw2910}

\bibitem[{Zhu {et~al.}(2025)Zhu, Acedo, Artuc, \& Chen}]{zhu2025}
Zhu, J., Acedo, E. d.~L., Artuc, K., \& Chen, X. 2025, RAS Techniques and
  Instruments, 4, rzae064

\bibitem[{{Zhu} {et~al.}(2021){Zhu}, {Su}, {Ji}, {Zhang}, {Zhao}, {Li}, {Dai},
  {Xue}, \& {Li}}]{Zhu2021}
{Zhu}, X.-Y., {Su}, Y., {Ji}, Y.-C., {et~al.} 2021, Research in Astronomy and
  Astrophysics, 21, 116, \dodoi{10.1088/1674-4527/21/5/116}

\bibitem[{Zonca {et~al.}(2019)Zonca, Singer, Lenz, Reinecke, Rosset, Hivon, \&
  Gorski}]{Zonca2019}
Zonca, A., Singer, L., Lenz, D., {et~al.} 2019, Journal of Open Source
  Software, 4, 1298, \dodoi{10.21105/joss.01298}

\end{thebibliography}
\bibliographystyle{aasjournal}

\end{document}